\documentclass{emulateapj}

\slugcomment{Astronomical Journal}

\shorttitle{L-band spectroscopy of IRLGs}
\shortauthors{Imanishi}

\begin{document}


\title{Infrared 3--4 $\mu$m Spectroscopy of Infrared Luminous Galaxies
with Possible Signatures of Obscured Active Galactic Nuclei} 


\author{Masatoshi Imanishi\altaffilmark{1,2,3}}
\affil{National Astronomical Observatory, 2-21-1, Osawa, Mitaka, Tokyo
181-8588, Japan}
\email{masa.imanishi@nao.ac.jp}

\altaffiltext{1}{Based in part on data collected at Subaru Telescope,
which is operated by the National Astronomical Observatory of Japan.}

\altaffiltext{2}{Visiting Astronomer at the Infrared Telescope Facility,
which is operated by the University of Hawaii under
Cooperative Agreement no. NCC 5-538 with the National
Aeronautics and Space Administration, Office of Space
Science, Planetary Astronomy Program.}

\altaffiltext{3}{Department of Astronomy, School of Science, Graduate
University for Advanced Studies, Mitaka, Tokyo 181-8588}

\begin{abstract}
We present the results of infrared 2.8--4.1 $\mu$m ($L$-band)
spectroscopy of nearby infrared luminous galaxies with possible
signatures of dust-obscured active galactic nuclei (AGNs) in data at
other wavelengths.  
The samples are chosen to include sources with a radio excess relative
to far-infrared emission, strong absorption features in mid-infrared
5--11.5 $\mu$m spectra, unusually weak [CII] 158 $\mu$m emission
relative to the far-infrared continuum, and radio galaxies classified
optically as narrow-line objects. Our aim is to investigate whether
the signatures of possible obscured AGNs can be detected in our $L$-band
spectra, based on the strengths of emission and absorption features.
Six of nine observed sources clearly show 3.3 $\mu$m polycyclic
aromatic hydrocarbon (PAH) emission features, a good starburst
indicator. An absorption feature at 3.1 $\mu$m due to ice-covered dust
is detected in IRAS 04154+1755 and IRAS 17208$-$0014. 
The signature of a bare carbonaceous dust absorption feature at 3.4
$\mu$m is seen in NGC 1377. Our $L$-band spectra reveal strong signatures
of obscured AGNs in all three optical Seyfert 2 galaxies (IRAS
04154+1755, Cygnus A, and 3C 234), and two galaxies classified
optically as non-Seyferts (NGC 828 and NGC 1377).
Among the remaining optical non-Seyferts, 
IRAS 17208$-$0014 might also show a buried AGN signature, whereas no
explicit AGN evidence is seen in the $L$-band spectra of the 
mid-infrared absorption-feature source IRAS 15250+3609, and two weak
[CII] emitters IC 860 and CGCG 1510.8+0725.  
\end{abstract}

\keywords{galaxies: active --- galaxies: ISM --- galaxies: nuclei ---
galaxies: Seyfert --- galaxies: starburst --- infrared: galaxies --- 
galaxies: individual: IRAS 04154+1755,  NGC 828, IRAS 15250+3609, 
IRAS 17208$-$0014, NGC 1377, IC 860, CGCG 1510.8+0725, Cygnus A, 3C 234}

\section{Introduction}

Infrared luminous galaxies (L$_{\rm IR}$ $\gtrsim$
10$^{11}$L$_{\odot}$) radiate most of their luminosity as dust
emission in the infrared \citep{sam96}. Thus, powerful energy sources,
e.g., starbursts and/or active galactic nuclei (AGNs), must be present
hidden behind dust. In the local universe, these infrared luminous
galaxies dominate the bright end of the luminosity function 
\citep{soi87}. In the distant universe, they dominate the cosmic
infrared background emission and have been used to trace the
dust-obscured star-formation rate, dust content, and metallicity in
the early universe, based on the assumption that they are powered by
starbursts \citep{bar00}. Estimating the AGN contribution to their
infrared luminosities is therefore important, not only to obtain a full
understanding of the nature of these galaxies, but also to study the
connections between obscured AGNs and starbursts in the universe.

In a powerful AGN that is obscured along our sightline by dust in a
{\it torus} geometry, the hard radiation of AGN can photo-ionize clouds
along the torus axis, above the torus scale height, and can produce the
so-called narrow line regions (NLRs; Antonucci 1993; Robson 1996). 
Since the NLRs show 
optical spectra that differ from normal star-forming galaxies, such
obscured AGNs should be detectable through optical spectroscopy
(classified as Seyfert 2s; Veilleux \& Osterbrock 1987). However,
since the nuclear regions of infrared luminous galaxies are very dusty
\citep{sam96}, virtually all lines of sight can be opaque to the bulk
of the ionizing radiation of AGNs. In such {\it buried} AGNs 
%-----------
\footnote{
We use the term ``obscured AGN'' if the AGN is obscured along
our line of sight, and ``buried AGN'' if the AGN is obscured along
virtually all sightlines.
}, 
%-----------
X-ray dissociation regions (XDRs), which are dominated by low-ionization
species \citep{mal96}, develop instead of the NLRs, so that it is
difficult to find AGN signatures using either conventional optical
spectroscopy or high-resolution infrared spectroscopic searches for
high-excitation forbidden emission lines that originate in the NLRs
\citep{gen98,vei99,mur01}. From the spectral shape of the cosmic X-ray
background emission, it is inferred that buried AGNs are abundant in
the universe \citep{fab02}; therefore, establishing a method with which
to detect such optically elusive \citep{mai03} buried AGNs is very
important.

To find such buried AGNs, it is necessary to conduct observations at
wavelengths with low dust extinction. Infrared $L$-band (2.8--4.1 $\mu$m)
spectroscopy from the ground is one of the most powerful tools for
this purpose. First, dust extinction in the $L$-band is lower than at
shorter wavelengths (A$_{\rm L}$ $\sim$ 0.06 $\times$ A$_{\rm V}$;
Rieke \& Lebofsky 1985) and is as low as at 5--13 $\mu$m
\citep{lut96}. Second, the 
contribution from AGN-powered dust emission to an observed flux,
relative to stellar emission, increases substantially
compared to $\lambda <$ 2 $\mu$m. Therefore, the detection of buried
AGNs becomes more feasible than at $\lambda <$ 2 $\mu$m. Third,
emission from a normal starburst and AGN are clearly
distinguishable based on $L$-band spectra \citep{moo86,imd00}. A
strong polycyclic aromatic hydrocarbon (PAH) emission feature is
usually found at 3.3 $\mu$m in a normal starburst galaxy
\citep{moo86}. In a pure normal starburst, high equivalent-width 3.3
$\mu$m PAH emission (EW$_{\rm 3.3PAH}$ $\sim$ 100 nm; Moorwood 1986;
Imanishi \& Dudley 2000) should always be found, since the equivalent
width is, by definition, insensitive to dust extinction of the
starburst. In a pure AGN, PAH molecules are destroyed, rather than excited, 
by X-ray radiation of the AGN \citep{voi92,sie04}, so that no 3.3
$\mu$m PAH emission is expected \citep{moo86}. Instead, a featureless
continuum from submicron-sized \citep{dra84,mat89} hot dust heated by
an AGN is found, making a PAH-free smooth 3--4 $\mu$m spectrum. In an
AGN/starburst composite galaxy, 3.3 $\mu$m PAH emission from the
starburst should be found, unless the starburst regions are directly
exposed to the X-ray radiation of the AGN; however, the PAH emission is
strongly diluted by the AGN-originating PAH-free continuum, so the
EW$_{\rm 3.3PAH}$ value should be reduced compared to a pure normal
starburst galaxy. Thus, the presence of an obscured AGN can be
established using the EW$_{\rm 3.3PAH}$ value.

If a highly dust-obscured AGN is present, and yet the AGN emission
contributes significantly to the observed $L$-band flux, then
absorption features at 3.1 and 3.4 $\mu$m produced by ice-covered dust
and bare carbonaceous dust, respectively, should be found
\citep{imd00,imm03,ris03,ris06,idm06}. Although a normal starburst,
in which stellar energy sources and dust are spatially well mixed
\citep{pux91,mcl93,for01}, can also show these absorption features
\citep{stu00}, there is an upper limit for the absorption optical
depths \citep{imm03,idm06}. This is because in this mixed source/dust
geometry, the less-obscured, less-attenuated emission from the
foreground (which shows weak dust absorption features) makes a
stronger contribution to the observed flux than the emission from the
distant side (which shows strong dust absorption features but is
highly attenuated). Thus, strong dust absorption features whose
optical depths are larger than the upper limit determined by the mixed
source/dust geometry indicate that obscuring dust is present as a
foreground screen distribution \citep{imm03,idm06}. A compact energy
source that is more centrally concentrated than the surrounding dust,
such as a buried AGN, is a natural explanation \citep{imm03,idm06},
unless a large amount of dust in a nearly edge-on host galaxy is
present in front of the primary 3--4 $\mu$m continuum-emitting energy
source.

\citet{idm06} presented $L$-band spectra of a large number of nearby
ultraluminous infrared galaxies (L$_{\rm IR}$ $>$ 10$^{12}$L$_{\odot}$;
Sanders \& Mirabel 1996) in the well-studied {\it IRAS} 1 Jy sample
\citep{kim98}, searched for possible signatures of powerful buried AGNs
in these galaxies, and statistically investigated the properties of
buried AGNs.  
However, there also exist several ultraluminous infrared galaxies 
(L$_{\rm IR}$ $>$ 10$^{12}$L$_{\odot}$) not included in the {\it IRAS} 
1 Jy sample, as well as galaxies with L$_{\rm IR}$ $<$
10$^{12}$L$_{\odot}$, that show possible signatures of AGNs obscured
behind dust in data at other wavelengths. 
In this paper, we present ground-based $L$-band spectra of
such galaxies, with the aim of testing if this $L$-band spectroscopic
method is indeed effective at finding AGN signatures.
Throughout this paper, $H_{0}$ $=$ 75 km s$^{-1}$ Mpc$^{-1}$, 
$\Omega_{\rm M}$ = 0.3, and $\Omega_{\rm \Lambda}$ = 0.7 are adopted.

\section{Targets}

Infrared luminous galaxies that have the following properties were
observed. 
(1) They do not belong to ultraluminous infrared galaxies in the 
{\it IRAS} 1 Jy sample and thus were not presented by \citet{idm06}.  
(2) Observations at other wavelengths have indicated at least some
signatures of obscured AGNs.  
(3) No high-quality $L$-band spectra were published in the
literature at the time of our observations.  
Table 1 summarizes the observed targets. 
The sample selection is heterogeneous, but our $L$-band spectra
can provide useful information about the nature of these interesting
galaxies. 

\subsection{Radio excess relative to far-infrared emission}

In normal starburst galaxies, the radio and far-infrared luminosities
are correlated, with a relatively small scatter \citep{con91}. 
If a galaxy shows a 
significant radio excess above this correlation, then the presence of
a radio-intermediate (or radio-loud) AGN is suggested. \citet{cra96}
investigated the radio to far-infrared luminosity ratios of
ultraluminous infrared galaxies and found that IRAS 04154+1755 shows 
(1) the largest radio excess and (2) a jet-like radio structure.
This object is classified optically as a Seyfert 2 \citep{cra96}, so
that a powerful radio-intermediate AGN must be present, hidden behind
torus-shaped dust. No significant 2--10 keV X-ray emission from the
putative AGN was detected \citep{imu99}, which suggests that the AGN
suffers from Compton-thick (N$_{\rm H}$ $>$ 10$^{24}$ cm$^{-1}$) X-ray
absorption.

\subsection{Strong ice absorption in mid-infrared 5--11.5 $\mu$m spectra}

Mid-infrared 5--11.5 $\mu$m spectra of many infrared luminous galaxies
were taken with {\it ISO}. From these spectra, \citet{spo02} selected
18 sources that show strong ice absorption features at 6.0 $\mu$m
(their Table 1, excluding NGC 253 and M82). To detect ice absorption,
a significant quantity of dust grains covered with an
ice mantle must be present in the foreground of the continuum-emitting
source(s). Such ice-covered dust grains are usually found deep inside
molecular gas, where dust is sufficiently shielded from ambient UV
radiation \citep{whi88}. A normal starburst galaxy with mixed
source/dust geometry can produce ice absorption features, if a
sufficient quantity of ice-covered dust is present. An obscured AGN is
an alternative candidate, particularly for sources with strong
absorption and weak PAH emission features (see $\S$1).

In the 5--11.5 $\mu$m spectra, PAH emission features are found at 6.2,
7.7, and 8.6 $\mu$m. In addition to the H$_{2}$O ice absorption at 6.0
$\mu$m, there are also absorption features due to hydrogenated
amorphous carbon (HAC) at 6.85 and 7.3 $\mu$m, and due to silicate
dust at 9.7 $\mu$m. Since these emission and absorption features
overlap significantly in wavelength, it is often difficult to
distinguish whether these galaxies are absorption-dominated (obscured
AGN candidates) or PAH-emission-dominated (normal starbursts)
\citep{spo02}. $L$-band spectroscopy may help to distinguish
whether the sources are dominated by PAH emission or absorption
features \citep{ima00a,ima00b,imd00}.

Among the 18 strong ice absorption sources, eight (IRAS
00188$-$0856, IRAS 05189$-$2524, UGC 5101, NGC 4418, Mrk 231, NGC
4945, Mrk 273, Arp 220) have available high quality $L$-band spectra
\citep{spo00,imd00,idm01,imm03,ima04,idm06}. Of these eight sources,
$L$-band spectra of five sources display signatures of obscured AGNs
(IRAS 00188$-$0856, IRAS 05189$-$2524, UGC 5101, Mrk 231, and Mrk 273)
\citep{imd00,idm01,imm03,idm06}, suggesting that this sample contains
many obscured AGNs. For the remaining ten unobserved sources, we
exclude four southern sources with declinations $<$ $-$30$^{\circ}$
from our target list, because they are difficult to observe from Mauna
Kea, Hawaii, our main observation site. Of the remaining six sources, we
observed NGC 828, IRAS 15250+3609, and IRAS
17208$-$0014, of which IRAS 15250+3609 shows the strongest absorption
and weakest PAH emission features, making this source the strongest
obscured AGN candidate. In addition, NGC 1377 (IRAS 03344$-$2103) is
also included, because its mid-infrared spectrum \citep{lau00} is
interpreted as dominated by strong 9.7 $\mu$m silicate dust absorption
($\tau_{9.7}$ $>$ 3.5), with weak PAH emission \citep{spo04}.

\subsection{Weak [CII] 158 $\mu$m emission relative to the far-infrared
continuum} 

\citet{mal97} measured the [CII] 158 $\mu$m emission lines of 30
optically normal (non-Seyfert) star-forming galaxies, and found that
three sources, NGC 4418, IC 860, and CGCG 1510.8+0725, show a significant
[CII] deficiency relative to their far-infrared continuum
luminosities. The presence of buried AGNs is a possible explanation
for the [CII] deficiency \citep{mal97}, although alternative scenarios
have also been proposed \citep{mal97,mal01}. In fact, detailed infrared
spectroscopic studies of NGC 4418 have provided many pieces of
evidence for a powerful buried AGN \citep{dud97,spo01,ima04}. For this
reason, IC 860 and CGCG 1510.8+0725 are also included as targets, in
the expectation that they may be objects similar to NGC 4418. 

\subsection{Narrow-line radio galaxies}

Narrow-line radio galaxies show Seyfert-2-type optical spectra.
According to the AGN unification paradigm, these galaxies are taken 
to contain powerful radio-loud AGNs hidden behind torus-geometry dust
\citep{urr95}. It is of interest to investigate whether our $L$-band
spectroscopic method can indeed successfully detect clear signatures
of obscured AGNs in these galaxies that are already known to possess
obscured AGNs. Two nearby narrow-line radio galaxies, Cygnus A and 3C
234, are selected.
 
Cygnus A is a nearby well-studied narrow-line radio galaxy
\citep{ost75}. Strong, but moderately absorbed, power-law X-ray
emission was detected \citep{uen94,you02}, suggesting the presence
of an obscured AGN. Near-infrared (1--4 $\mu$m) observations also
suggested the presence of a powerful obscured AGN (A$_{\rm V}$ = 40--150
mag) \citep{djo94,war96,pac98,tad99}. Based on a mid-infrared 8--13 $\mu$m
spectrum, \citet{imu00} suggested that the energy source is more
centrally concentrated than the nuclear obscuring dust, as is expected
for an obscured AGN.

3C 234 shows a broad H$\alpha$ emission line in its optical spectrum
\citep{gra78}. 
However, \citet{ant90} and \citet{you98} argued that the broad
component is interpreted as a scattered component, rather than a
directly transmitted component, and that optical classification of
this galaxy as a narrow-line radio galaxy is more appropriate. The
dust extinction toward the AGN is estimated to be A$_{\rm V}$ = 60 mag
\citep{you98}. \citet{sam99} detected intrinsically luminous,
moderately absorbed X-ray emission from the putative AGN.

\section{Observations and Data Analysis}

Infrared $L$-band spectroscopic observations were made using 
infrared spectrographs, IRCS \citep{kob00} on the Subaru 8.2-m telescope
\citep{iye04} and SpeX \citep{ray03} on the IRTF 3-m telescope. 
Table 2 summarizes a detailed observation log. 

For Subaru IRCS observation runs, the sky was clear during the
observations and seeing sizes at $\lambda$ = 2.2 $\mu$m were measured
to be 0$\farcs$5--0$\farcs$8 in full-width at half maximum (FWHM). 
A 0$\farcs$9-wide slit and the $L$-grism were used with a 58-mas pixel
scale. The achievable spectral resolution was $\sim$140 at 3.5 $\mu$m. 
The position angle of the slit was set along the north-south direction.
A standard telescope nodding technique (ABBA pattern) with a throw of 7
arcsec along the slit was employed to subtract background emission.
Each exposure was 1.5--2.0 sec, and 30--40 coadds were made at each
position. 

For the IRTF SpeX observations, we employed the 1.9--4.2 $\mu$m
cross-dispersed mode with a 1\farcs6 wide slit. This mode enables $L$- 
(2.8--4.1 $\mu$m) and $K$-band (2--2.5 $\mu$m) spectra to be obtained
simultaneously, with a spectral resolution of R $\sim$ 500. The sky
conditions were clear throughout the observations, and seeing sizes at
$\lambda$ = 2.2 $\mu$m were in the range 0$\farcs$4--1$\farcs$0 FWHM.
The position angle of the slit was set along the north-south
direction. A standard telescope nodding technique with a throw of 7.5
arcsec was employed along the slit. The telescope tracking was
monitored with the infrared slit-viewer of SpeX. Each exposure was 15
sec, and 2 coadds were made at each position.

A-, F-, and G-type main sequence stars (Table~\ref{tbl-2}) were
observed as standard stars, with airmass differences of $<$0.1 to the
individual target objects, to correct for the transmission of the
Earth's atmosphere. The $K$- and $L$-band magnitudes of the standard
stars were estimated from their $V$-band (0.6 $\mu$m) magnitudes,
adopting the $V-K$ and $V-L$ colors appropriate to the stellar types
of individual standard stars, respectively \citep{tok00}.

Standard data analysis procedures were employed, using the Image Reduction 
and Analysis Facility (IRAF) 
%--------------
\footnote{
IRAF is a general purpose software system distributed by the National
Optical Astronomy Observatories, which are operated by the Association
of Universities for Research in Astronomy, Inc. (AURA), under
cooperative agreement with the  National Science Foundation.}. 
%--------------
Initially, frames taken with an A (or B) beam were subtracted from
frames taken subsequently with a B (or A) beam, and the resulting
subtracted frames were added and divided by a spectroscopic flat
frame.  Bad pixels and pixels hit by cosmic rays were then replaced
with the interpolated values from surrounding pixels.  Corrections for
cosmic ray hits were much larger for IRTF SpeX than Subaru IRCS.
Finally, the spectra of the target objects and standard stars were
extracted.  Wavelength calibration was performed taking into account
the wavelength-dependent transmission of the Earth's atmosphere.  The
spectra of the targets were divided by the observed spectra of
standard stars, multiplied by the spectra of blackbodies with
temperatures appropriate to the individual standard stars
(Table~\ref{tbl-2}), and then flux-calibrated. Appropriate binning of
spectral elements was performed to achieve an adequate signal-to-noise
ratio ($\gtrsim$10) in most of the elements, particularly at 
$\lambda_{\rm obs} < 3.3$ $\mu$m in the observed frame, where the
Earth's atmospheric transmission is poorer than at longer $L$-band
wavelengths. 
For all sources, whole spectral datasets were divided into a few 
sub-groups, and error bars at each spectral element were estimated
from the scatter of actual signals among the sub-groups. 

\section{Results}

\subsection{$L$-band spectra}

\subsubsection{3.3 $\mu$m PAH emission}

Figure 1 shows flux-calibrated nuclear $L$-band spectra of the nine
observed sources. All sources, except NGC 1377, Cygnus A, and
3C 234, show clear emission features at $\lambda_{\rm obs}$ $\sim$ (1
+ $z$) $\times$ 3.3 $\mu$m, which we identify as the 3.3 $\mu$m PAH
emission. The detection of the 3.3 $\mu$m PAH emission, an indicator
of starbursts, means that these galaxies contain detectable starburst
activity. To estimate the 3.3 $\mu$m PAH flux, we adopt a template
spectral shape for Galactic star-forming regions and nearby starburst
galaxies (type-1 sources; Tokunaga et al. 1991). In this template, the
main 3.3 $\mu$m PAH emission profile extends between $\lambda_{\rm
rest}$ $=$ 3.24--3.35 $\mu$m in the rest frame. Data
points at wavelengths slightly shorter than $\lambda_{\rm rest}$ $=$
3.24 $\mu$m and slightly longer than $\lambda_{\rm rest}$ $=$ 3.35
$\mu$m, unaffected by obvious absorption features, are adopted as the
continuum levels, which are shown as solid lines for PAH-detected
sources in Figure 1. 
For sources with clearly detectable PAH emission, this template profile
reproduces the observed 3.3 $\mu$m PAH emission features reasonably
well, with our spectral resolution and signal-to-noise ratios, except 
NGC 828 which displays a slightly broader profile than the template  
(Figure 1, dotted lines).
Table 3 summarizes the observed properties of the 3.3 $\mu$m PAH
emission features.  
The uncertainties of the 3.3 $\mu$m PAH fluxes, estimated from
the fittings, are $<$20\% in all cases. 
Possible systematic uncertainties coming from continuum determination
ambiguities are also unlikely to exceed 20\%, as long as reasonable
continuum levels are adopted.  
Thus, total uncertainties of the PAH fluxes are taken as $<$30\%. 
Cygnus A, 3C 234, and NGC 1377 may also show excesses at the expected
wavelength of 3.3 $\mu$m PAH emission, but their significance is
marginal; we therefore estimate upper limits for the PAH fluxes.

\subsubsection{3.1 $\mu$m H$_{2}$O ice absorption}

IRAS 04154+1755 ($z =$ 0.056) shows a concave continuum, which is
naturally explained by the strong, broad 3.1 $\mu$m H$_{2}$O ice
absorption feature \citep{spo00,imm03,idm06}. The detection of this
ice absorption feature means that the $L$-band continuum-emitting
energy source is obscured by ice-covered dust grains deep inside
molecular gas \citep{whi88}. Following \citet{idm06}, we adopt a
linear continuum to estimate the optical depth of this 3.1 $\mu$m
H$_{2}$O ice absorption ($\tau_{3.1}$). The continuum level varies
slightly depending on the adopted data points used for the continuum
determination. We adopt a relatively low continuum level to obtain
a conservative lower limit for the $\tau_{3.1}$ value in IRAS
04154+1755.
It is $\tau_{3.1}$ $\sim$ 0.9, which corresponds to A$_{\rm
V}$ $\sim$ 15 mag if the Galactic $\tau_{3.1}$/A$_{\rm V}$ ratio
($\sim$0.06; Smith et al. 1993; Tanaka et al. 1990; Murakawa et al.
2000) is assumed. The ice absorption feature of IRAS 04154+1755 has a
strong absorption wing at $\lambda_{\rm obs}$ = 3.6--3.9 $\mu$m or
$\lambda_{\rm rest}$ = 3.4--3.7 $\mu$m, which resembles the absorption
profile of GL 2591 (a Galactic protostar surrounded by circumstellar
material), rather than Elias 16 (a Galactic late-type field star
behind molecular clouds) \citep{smi89}.

The continuum of IRAS 17208$-$0014 ($z =$ 0.042) is also concave-shaped.
We obtain $\tau_{3.1}$ $\sim$ 0.4, which is similar to the recent
estimate by \citet{ris06}, using independent data.
Unlike IRAS 04154+1755, the ice absorption feature of IRAS 17208$-$0014
has a relatively weak absorption wing at $\lambda_{\rm obs}$ = 3.55--3.85
$\mu$m or $\lambda_{\rm rest}$ = 3.4--3.7 $\mu$m, as seen in Elias 16,
rather than GL 2591 \citep{smi89}. 

\subsubsection{3.4 $\mu$m bare carbonaceous dust absorption}

NGC 1377 ($z =$ 0.005) shows an absorption feature that peaks at
$\lambda_{\rm obs}$ = 3.4--3.45 $\mu$m, or $\lambda_{\rm rest}$ $\sim$
3.4 $\mu$m. We interpret this to be the 3.4 $\mu$m dust absorption
feature produced by bare carbonaceous dust grains, which is found in
Galactic stars highly reddened by the {\it diffuse} interstellar
medium outside molecular gas \citep{pen94,ima96,raw03}. Its optical
depth $\tau_{3.4}$ is estimated to be 0.17, which corresponds to
A$_{\rm V}$ = 25--40 mag, if the dust extinction curve in NGC 1377 is
similar to the Galactic diffuse interstellar medium
($\tau_{3.4}$/A$_{\rm V}$ $\sim$ 0.004--0.007; Pendleton et al. 1994).

For Cygnus A ($z =$ 0.056) and 3C 234 ($z =$ 0.185), the presence of the
3.4 $\mu$m absorption feature in our $L$-band spectra is uncertain. We
estimate conservative upper limits of $\tau_{3.4}$ $<$ 0.2 for both
sources, which corresponds to A$_{\rm V}$ $<$ 50 mag toward the $L$-band
continuum emitting regions, if the Galactic relation of
$\tau_{3.4}$/A$_{\rm V}$ 
$\sim$ 0.004--0.007 is assumed. The A$_{\rm V}$ value derived from our
$L$-band spectra is at the lower end of the range for Cygnus A
(A$_{\rm V}$ = 40--150 mag; $\S$2.4) and is smaller than the other
estimate for 3C 234 (A$_{\rm V}$ = 60 mag; $\S$2.4).

\subsubsection{[MgVIII]3.028$\mu$m emission}

3C 234 ($z =$ 0.185) shows an emission line at $\lambda_{\rm obs}$
$\sim$ 3.58 $\mu$m or $\lambda_{\rm rest}$ $\sim$ 3.02 $\mu$m, which
we identify as the [Mg VIII] 3.028$\mu$m line. We estimate its flux to
be f([MgVIII]) = 1.5 $\times$ 10$^{-14}$ ergs s$^{-1}$ cm$^{-2}$, and
its luminosity to be L([MgVIII]) = 1.2 $\times$ 10$^{42}$ ergs
s$^{-1}$.

\subsection{$K$-band spectra}

For the two sources observed with IRTF SpeX, i.e., IC 860 and CGCG
1510.8+0725 (weak [CII] emitters), $K$-band spectra are obtained
simultaneously with the $L$-band spectra. Figure 2 shows these 
$K$-band spectra.
 
The spectra of IC 860 ($z =$ 0.013) and CGCG 1510.8+0725 ($z =$ 0.013)
show gaps in the continuum at $\lambda_{\rm obs} \sim$ 2.35 $\mu$m. We
attribute these gaps to CO absorption features at $\lambda_{\rm rest}$ =
2.31--2.4 $\mu$m produced by stars \citep{iva00,ia04,imw04}. For the
other weak [CII] 
emitter, NGC 4418 \citep{mal97}, the CO absorption feature is clearly
detected \citep{ima04}. To estimate the CO absorption strengths in IC
860 and CGCG 1510.8+0725, we adopt the spectroscopic CO index
(CO$_{\rm spec}$) defined by \citet{doy94} and follow the procedures
applied to NGC 4418 \citep{ima04}. For both IC 860 and CGCG
1510.8+0725, power-law continuum levels (F$_{\rm \lambda}$ = $\alpha
\times \lambda^{\beta}$) are determined using data points at
$\lambda_{\rm obs}$ = 2.08--2.32 $\mu$m ($\lambda_{\rm rest}$ =
2.05--2.29 $\mu$m), excluding obvious emission lines. The adopted
continuum levels are shown as dashed lines in Figure 2. Data at
$\lambda_{\rm obs}$ = 2.34--2.43 $\mu$m ($\lambda_{\rm rest}$ =
2.31--2.40 $\mu$m) are used to derive the CO$_{\rm spec}$ values. We
obtain CO$_{\rm spec}$ $\sim$ 0.1 for IC 860 and $\sim$0.2 for CGCG
1510.8+0725.
However, the continuum determination of IC 860 is less secure than 
CGCG 1510.8+0725 because of the large scatter of continuum data points 
in the former (Figure 2). 
We try several continuum levels within a reasonable range, and 
obtain the largest plausible value of CO$_{\rm spec}$ $\sim$ 0.2 for IC 860.

The typical values observed in star-forming or elliptical (=spheroidal)  
galaxies are CO$_{\rm spec}$ $>$ 0.15 \citep{gol97a,gol97b,iva00}.
Dilution of the CO absorption feature by a featureless continuum 
from AGN-heated hot dust or from stars younger than a few
million years \citep{lei99} can decrease the CO$_{\rm spec}$ values.
Based on CO$_{\rm spec}$ values, there is no explicit AGN evidence in 
IC 860 and CGCG 1510.8+0725.

The $K$-band spectrum of CGCG 1510.8+0725 shows strong H$_{2}$
emission features, as with NGC 4418 \citep{ima04}. For NGC 4418,
strong H$_{2}$ emission lines and other observational properties are
naturally explained by a powerful buried AGN
\citep{dud97,spo01,ima04}. 
However, the signal-to-noise ratios for CGCG 1510.8+0725 are not good
enough to investigate in a quantitatively detailed manner whether
the strong H$_{2}$ emission originates in a buried AGN or other
mechanisms related to starbursts. 

\section{Discussion}

\subsection{Detected modestly obscured starbursts}

3.3 $\mu$m PAH emission, a signature of starbursts, is detected in six
of nine observed infrared luminous galaxies. Since dust extinction
in the $L$-band is about 0.06 times as large as
that in the optical $V$-band ($\lambda$ = 0.6 $\mu$m; Rieke \& Lebofsky
1985), the flux of the 3.3 $\mu$m PAH emission is not significantly 
attenuated ($<$1 mag) if dust extinction is less than 15 mag in
A$_{\rm V}$. Thus, the observed 3.3 $\mu$m PAH emission luminosities
are a good measure of the absolute luminosity of modestly obscured
(A$_{\rm V}$ $<$ 15 mag) starburst activity.

Table 3 summarizes the observed 3.3 $\mu$m PAH to infrared luminosity
ratios (L$_{\rm 3.3PAH}$/L$_{\rm IR}$). The ratios are factors of 3 to
$>$10 lower than those reported for modestly obscured starbursts 
($\sim$10$^{-3}$; Mouri et al. 1990; Imanishi 2002). If these
ratios are taken at face value, the detected modestly obscured
starbursts can account for only some fraction of the infrared
luminosities of these galaxies. The infrared luminosities must
therefore be dominated by AGNs and/or very heavily obscured 
(A$_{\rm V}$ $>>$ 15 mag) starbursts. 

\subsection{AGNs obscured by torus-shaped dust in optical Seyfert 2s}

IRAS 04154+1755, Cygnus A, and 3C 234 are classified optically as
Seyfert 2s (Table 1). Their optical Seyfert 2 classifications indicate
that (1) the AGNs are obscured along our line of sight by torus-shaped
dust, (2) a large amount of the AGNs' ionizing photons are escaping
along the torus axis, and (3) emission lines, originating in the
well-developed narrow line regions (NLRs), are strong and clearly
detectable in the optical spectra. For these optical Seyfert 2s, the
presence of obscured AGNs is much more certain than in other infrared
luminous galaxies with no optical Seyfert signatures (non-Seyferts).
Hence, it is useful to test whether our $L$-band spectroscopic method
can succeed in detecting AGN signatures in these optical Seyfert 2s.

For Cygnus A and 3C 234, the 3.3 $\mu$m PAH equivalent widths (Table
3) are more than an order of magnitude lower than those of typical
starburst galaxies (EW$_{\rm 3.3PAH}$ $\sim$100 nm; Moorwood 1986;
Imanishi \& Dudley 2000). The observed $L$-band fluxes must be
dominated by PAH-free emission, most plausibly a featureless continuum
from AGN-heated hot dust. Thus, our $L$-band spectroscopy has
successfully revealed AGN signatures in these two optical Seyfert 2s.

For 3C 234, the [Mg VIII]3.028$\mu$m emission line is detected. The
ionization potential of the [Mg VIII] line is as high as 224.9 eV, and
this emission line has been detected in several nearby Seyfert
galaxies \citep{lut00,stu02,mar03}. The detection of such a high
excitation forbidden emission line requires a very hard radiation
field, further supporting the presence of an AGN.

The [Mg VIII] luminosity of 3C 234 is larger than the
prototypical nearby Seyfert 2 galaxy NGC 1068 by a factor of 40 \citep{mar03}. The observed 2--10 keV X-ray luminosity from the AGN in NGC 1068 is
estimated to be $\sim$1 $\times$ 10$^{41}$ ergs s$^{-1}$, but this is
believed to be a reflected component \citep{koy89}. Assuming a
reflected fraction of 0.01 \citep{pie94}, the intrinsic 2--10 keV
luminosity in NGC 1068 is Lx(2--10 keV) $\sim$ 10$^{43}$ ergs
s$^{-1}$. If the [Mg VIII] to 2--10 keV luminosity ratio in 3C 234 is
similar to that of NGC 1068, then the predicted 2--10 keV luminosity
for 3C 234 is Lx(2--10 keV) $\sim$ 4 $\times$ 10$^{44}$ ergs s$^{-1}$,
which is roughly comparable to the measured value of $\sim$1 $\times$
10$^{44}$ ergs s$^{-1}$ \citep{sam99}.

For IRAS 04154+1755, the presence of an AGN is less certain than for
Cygnus A and 3C 234, if we consider the EW$_{\rm 3.3PAH}$ value
($\sim$60 nm for IRAS 04154+1755; Table 3). However, the relatively
large EW$_{\rm 3.3PAH}$ value in IRAS 04154+1755 is probably the
consequence of the strong ($\tau_{3.1}$ $\sim$ 0.9) 3.1 $\mu$m H$_{2}$O
ice absorption. This absorption feature strongly attenuates the
AGN-originated flux and reduces the dilution of the 3.3 $\mu$m PAH
emission by an AGN continuum, making the EW$_{\rm 3.3PAH}$ value
relatively large, even in the presence of a powerful obscured AGN. We
therefore search for signatures of an obscured AGN based on a
$\tau_{3.1}$ value. 

As mentioned in $\S$1, and in more detail in \citet{imm03} and
\citet{idm06}, $\tau_{3.1}$ values can be used to distinguish between
a normal starburst (mixed source/dust geometry, small $\tau_{3.1}$) or
an obscured AGN (centrally-concentrated energy source geometry, large
$\tau_{3.1}$). 
Since the fraction of dust covered with an ice mantle in the majority of
infrared luminous galaxies is likely to be smaller than the well-studied
less infrared-luminous starburst galaxy M82, because of harsher
radiation environment in the former \citep{soi00,soi01}, $\tau_{3.1}$
larger than 0.3 suggests a centrally-concentrated energy source geometry
\citep{imm03,idm06}.  
The large observed $\tau_{3.1}$ of IRAS 04154+1755 ($\sim$0.9) strongly
indicates a centrally-concentrated energy source, such as an obscured
AGN. We note 
that IRAS 04154+1755 is located close to the Galactic Taurus molecular
clouds, but no significant molecular gas is detected in the exact
direction of 
IRAS 04154+1755 \citep{ung87}. Furthermore, the wavelength range in which
H$_{2}$O ice absorption is strong in Figure 1 is similar
to that expected from the profile of GL 2591 redshifted with $z =$
0.056, rather than $z =$ 0. Thus, the strong 3.1 $\mu$m
H$_{2}$O absorption in IRAS 04154+1755 is believed to be nuclear
in origin.

In summary, our $L$-band spectroscopic method successfully provides
AGN evidence for the three optical Seyfert 2s, which are known to
possess luminous AGNs behind torus-shaped dust. \citet{idm06} also
detected AGN signatures in the $L$-band spectra of all observed
ultraluminous infrared galaxies classified optically as Seyferts. 
We can therefore conclude 
that our $L$-band spectroscopic method is at least as effective as 
conventional optical spectroscopy for the purpose of finding AGNs
obscured behind dust. However, the question of how much more effective
our $L$-band method is depends on the detection rate of buried AGNs in
optical non-Seyfert galaxies.

\subsection{Buried AGNs in NGC 828 and NGC 1377}

Among the six observed optical non-Seyfert galaxies (NGC 828, IRAS
15250+3609, IRAS 17208$-$0014, NGC 1377, IC 860, and CGCG 1510.8+0725),
the EW$_{\rm 3.3PAH}$ values of NGC 828 and NGC 1377 are $\lesssim$ 20
nm. Starburst galaxies have an average value of EW$_{\rm 3.3PAH}$ $\sim$
100 nm, with some scatter, 
but never lower than $\sim$40 nm \citep{moo86}. The small EW$_{\rm
3.3PAH}$ values of NGC 828 and NGC 1377 suggest that powerful
AGNs are present, which heat up dust grains and produce strong PAH-free
featureless continua that dilute the 3.3 $\mu$m PAH
emission from starbursts. Our $L$-band spectroscopy thus newly reveals
buried AGN signatures in two optical non-Seyfert galaxies.

The $L$-band continuum of NGC 1377 is unusually red compared to normal
starburst emission. Such red $L$-band continua have previously been
observed in the ultraluminous infrared galaxies Superantennae \citep{ris03} 
and IRAS 08572+3915 \citep{idm06}, both of which show very weak 3.3
$\mu$m PAH emission and are therefore classified as
obscured-AGN-dominated. Very highly reddened hot dust emission heated
by an obscured AGN is the most natural explanation for the extremely
red $L$-band continuum of NGC 1377 \citep{ris06}. The observed
$\tau_{3.4}$ value of 0.17 corresponds to A$_{\rm V}$ = 25--40 mag,
or A$_{\rm 3.5 \mu m}$ = 1.4--2.5 mag if the Galactic dust extinction
curve is adopted. Thus, the dereddened luminosity heated by an AGN is
$\nu$L$_{\nu}$(3.8$\mu$m) = 0.4--1 $\times$ 10$^{43}$ ergs s$^{-1}$,
which can account for a significant fraction of the infrared
luminosity of NGC 1377 (L$_{\rm IR}$ $\sim$ 3 $\times$ 10$^{43}$ ergs
s$^{-1}$).

\subsection{Weak signatures of buried AGNs in IRAS 17208$-$0014 and CGCG
1510.8+0725}  

For the two optical non-Seyfert galaxies IRAS 17208$-$0014 and CGCG
1510.8+0725, some indications of buried AGNs might be present.

For IRAS 17208$-$0014, the large $\tau_{3.1}$ value of $\sim$0.4
suggests a centrally-concentrated energy source, such as a buried AGN,
if the fraction of ice-covered dust is comparable to or smaller than
that in the well-studied starburst galaxy M82.
However, \citet{soi00} found that the mid-infrared 12.5 $\mu$m 
dust emission from IRAS 17208$-$0014 is dominated by spatially
extended components, with no prominent core, indicating that dust
emission at this wavelength is dominated by extended starburst
activity. 
The estimated surface brightness of IRAS 17208$-$0014 is exceptionally
low compared to other ultraluminous infrared galaxies 
(L$_{\rm IR}$ $>$ 10$^{12}$L$_{\odot}$), and can even be smaller than
that of M82 \citep{soi00}.  
Hence, the fraction of ice-covered dust can be larger than that in M82, 
because of a weaker radiation density in IRAS 17208$-$0014. 
Unlike the majority of other infrared luminous galaxies, the observed
$\tau_{3.1}$ value of $\sim$0.4 in IRAS 17208$-$0014 could be reproduced
by a mixed source/dust geometry ($\S$5.2).   
Millimeter interferometric observations, based on the HCN (J = 1--0) to 
HCO$^{+}$ (J = 1--0) ratio in brightness temperature, have suggested
that a buried AGN might be present, but failed to reveal clear AGN
evidence \citep{ink06}.  
For this source, although absorption features at 5--11.5 $\mu$m are
found, 6.2 $\mu$m PAH emission is also strong \citep{spo02}. 
The presence of a luminous buried AGN in IRAS 17208$-$0014 has not been
explicitly indicated in any available data and so is currently unclear. 

For the weak [CII] emitter CGCG 1510.8+0725, strong H$_{2}$
emission might originate in a buried AGN ($\S$4.2), but the buried AGN
evidence is not strong.
Although the presence of a strong buried AGN is one possibility to
explain the small [CII] to far-infrared continuum luminosity ratio
\citep{mal97,mal01}, scenarios that do not invoke a powerful buried AGN,
such as large radiation density or optical depth effects in starbursts
\citep{mal97,mal01}, could be alternative possibilities for the weak 
[CII] emission. 
It is unclear whether CGCG 1510.8+0725 does in fact possess
a luminous buried AGN. 

\subsection{No explicit buried AGN signatures in IC 860 and IRAS 15250+3609} 

For the remaining two optical non-Seyfert galaxies IC 860 and IRAS 15250+3609,
no explicit buried AGN signatures are evident in our spectra. 
For the weak [CII] emitter IC 860, it is unclear whether 
there is indeed a powerful buried AGN, as in the case of CGCG
1510.8+0725 ($\S$5.4).

IRAS 15250+3609 shows strong absorption features, with very weak PAH
emission in its mid-infrared 5--11.5 $\mu$m spectrum. These
mid-infrared properties are very similar to NGC 1377, and yet the
$L$-band spectra are very different, in that only NGC 1377 shows 
strong buried AGN signatures. This discrepancy probably derives from the
different overall spectral energy distributions. Figure 3 shows
infrared 1--11.5 $\mu$m spectral energy distributions of NGC 1377 and
IRAS 15250+3609. The emission at 5--11.5 $\mu$m is dominated by dust
emission powered by obscured energy sources (starbursts and/or AGNs),
whereas that at 1--2.5 $\mu$m originates in foreground stellar
emission. The signatures of obscured AGNs can only be detected in the
$L$-band spectra if AGN-heated hot dust emission contributes
significantly to the observed $L$-band flux.

In Figure 3 ({\it Left}), the $L$-band emission of NGC 1377 is
smoothly connected to the longer wavelength dust emission spectrum
observed with {\it ISO}, which is judged to be powered by an obscured
AGN, because of strong absorption and lack of PAH emission features
\citep{lau00}. Detection of strong AGN signatures in the $L$-band
spectrum of NGC 1377 is possible primarily because the AGN-heated dust
emission is a dominant component at $L$. In contrast, the $L$-band
emission of IRAS 15250+3609 (Figure 3, {\it Right}) corresponds to a
transition between stellar and dust emission. The shorter part of the
$L$-band spectrum ($<$3.5 $\mu$m) is likely to be dominated by
foreground stellar emission, which is why we missed clear AGN signatures
in our $L$-band spectrum of IRAS 15250+3609, despite its 5--11.5
$\mu$m spectrum strongly suggesting the presence of a powerful obscured
AGN.

\section{Summary}

We reported infrared $L$-band spectroscopic results for nine infrared
luminous galaxies that may possess signatures of AGNs hidden
behind dust based on observational data at other wavelengths. A radio to
far-infrared excess galaxy (IRAS 04154+1755), four sources that
exhibit absorption features in the 5--11.5 $\mu$m spectra (NGC 828,
IRAS 15250+3609, IRAS 17208$-$0014, and NGC 1377), two weak [CII]
158$\mu$m emitters (IC 860 and CGCG 1510.8+0725), and two narrow-line
radio galaxies (Cygnus A and 3C 234) were observed. Using the 3.3
$\mu$m PAH emission feature, the [Mg VIII] 3.028 $\mu$m emission line,
and absorption features at 3.1 $\mu$m from ice-covered dust and at 3.4
$\mu$m from bare carbonaceous dust, we searched for signatures of
obscured AGNs in our $L$-band spectra. For the two weak [CII]
emitters, $K$-band spectra were also simultaneously taken. We found
the following main results:

\begin{enumerate}
\item For two optical Seyfert 2 galaxies (Cygnus A and 3C 234), our
      results strongly suggested that the observed $L$-band spectra are
      dominated by a PAH-free featureless continuum, originating in
      AGN-heated hot dust, because of very small 3.3 $\mu$m PAH
      equivalent widths. A strong detection of the [Mg VIII] 3.028
      $\mu$m line in the $L$-band spectrum of 3C 234 also supported
      the presence of a luminous AGN behind torus-shaped dust in this
      source.
\item For another optical Seyfert 2 galaxy, IRAS 04154+1755, an energy
      source that is more centrally concentrated than dust was
      suggested from the large optical depth of the 3.1 $\mu$m dust
      absorption feature ($\tau_{3.1}$ $\sim$ 0.9). An obscured AGN is
      a natural explanation for this energy source, because in
      normal starburst galaxies, stellar energy sources and dust are
      spatially well mixed.
\item Our $L$-band spectroscopic method succeeded in detecting clear
      AGN signatures in all three optical Seyfert 2s
      known to possess powerful AGNs behind torus-shaped dust.
      This means that as a tool for finding obscured AGNs, our $L$-band
      spectroscopic method is at least as powerful as conventionally used
      optical spectroscopy. 
\item Among the remaining six optical non-Seyferts (NGC 828, IRAS
      15250+3609, IRAS 17208$-$0014, NGC 1377, IC 860, and CGCG
      1510.8+0725), strong AGN signatures were found in NGC 828 and NGC 1377,
      based on their very small 3.3 $\mu$m PAH equivalent widths.
      NGC 1377 showed a very red $L$-band continuum, which also supports
      the obscured-AGN-dominated nature of this galaxy.
      The detection of AGNs in the two optical non-Seyferts clearly
      indicated that our $L$-band spectroscopic method is more effective
      than optical spectroscopy in finding obscured AGNs.
\item IRAS 17208$-$0014 and NGC 1377 may have shown 
      3.1 $\mu$m and 3.4 $\mu$m dust absorption features 
      with optical depths of $\tau_{3.1}$ $\sim$ 0.4 and $\tau_{3.4}$
      $\sim$ 0.17, respectively. 
      The large $\tau_{3.1}$ value of IRAS 17208$-$0014 indicated a 
      centrally-concentrated energy source, such as a buried AGN, if 
      the fraction of ice-covered dust is similar to or smaller than 
      that in the well-studied starburst galaxy M82.
\item Strong H$_{2}$ emission found in the $K$-band spectrum of the weak
      [CII] emitter CGCG 1510.8+0725 might come from phenomena related to
      a buried AGN, like another weak [CII] emitter NGC 4418. 
\item For IRAS 17208$-$0014 and CGCG 1510.8+0725, possible buried AGNs
      signatures were found, but weak.
      For IC 860 and IRAS 15250+3609, we failed to provide any explicit 
      buried AGN evidence in our spectra.
      The non-detection of buried AGN signatures in IRAS 15250+3609 was
      explained by its infrared 1--11.5 $\mu$m 
      spectral energy distribution, where foreground stellar emission is
      dominant at $L$. 
      However, for the remaining three optical non-Seyfert galaxies IRAS
      17208$-$0014, IC 860, and CGCG 1510.8+0725, the evidence for
      buried AGNs at other wavelengths is also very weak.
      The absence of buried AGN signatures in these sources may be due
      to their starburst-dominated nature. 
\end{enumerate}

We are grateful to M. Ishii, R. Potter, E. Pickett, A. Hatakeyama, M.
Lemmen, B. Golisch, and D. Griep for their support during our Subaru and
IRTF observations. 
We thank the anonymous referee for his/her useful comments.
Text data of ISO spectra of IRAS 15250+3609 and
NGC 1377 were kindly provided by H. W. W. Spoon. M.I. is supported by
Grants-in-Aid for Scientific Research (16740117). Some of the data
analysis was carried out using a computer system operated by the Astronomical
Data Analysis Center (ADAC) and the Subaru Telescope of the National
Astronomical Observatory, Japan. This research has made use of the
SIMBAD database, operated by the Centre de Donnees astronomiques de
Strasbourg (CDS), Strasbourg, France; of the NASA/IPAC
Extragalactic Database (NED), which is operated by the Jet Propulsion
Laboratory, California Institute of Technology, under contract with
the National Aeronautics and Space Administration (NASA); and of data
products from the Two Micron All Sky Survey, which is a joint project
between the University of Massachusetts and the Infrared Processing and
Analysis Center/California Institute of Technology, funded by NASA and
the National Science Foundation. 

%\clearpage

\clearpage

%%%%%%%%%% Table 1 %%%%%%%%%
\begin{deluxetable}{lcrrrrccc}
%\rotate
\tabletypesize{\scriptsize}
\tablecaption{Infrared Emission Properties of the Observed Infrared
Luminous Galaxies 
\label{tbl-1}}
\tablewidth{0pt}
\tablehead{
\colhead{Object} & \colhead{Redshift}   & 
\colhead{f$_{\rm 12}$}   & 
\colhead{f$_{\rm 25}$}   & 
\colhead{f$_{\rm 60}$}   & 
\colhead{f$_{\rm 100}$}  & 
\colhead{log $L_{\rm IR}$ (log $L_{\rm IR}$/$L_{\odot}$)} & 
\colhead{f$_{25}$/f$_{60}$} & 
\colhead{Optical}   \\
\colhead{} & \colhead{}   & \colhead{(Jy)} & \colhead{(Jy)} 
& \colhead{(Jy)} & \colhead{(Jy)}  & \colhead{(ergs s$^{-1}$)} & \colhead{}
& \colhead{Class}  \\
\colhead{(1)} & \colhead{(2)} & \colhead{(3)} & \colhead{(4)} & 
\colhead{(5)} & \colhead{(6)} & \colhead{(7)} & \colhead{(8)} & 
\colhead{(9)} 
}
\startdata
IRAS 04154+1755 & 0.056 & 0.20 & 0.71 & 3.82 & 5.84 & 45.4 (11.8) &
0.19 (C) &  Sy2 (i) \\  
NGC 828 & 0.018 & 0.70 & 1.03 & 10.9 & 22.8 & 44.9 (11.3) & 0.09 (C) &
HII (ii) \\  
IRAS 15250+3609 & 0.055 & $<$0.20 & 1.32 & 7.29 & 5.91 & 45.6 (12.0) & 
0.18 (C) & LINER (iii) \\   
IRAS 17208$-$0014 & 0.042 & 0.20 & 1.66 & 31.1 & 34.9 & 45.9 (12.3) &
0.05 (C) &  HII (iii) \\  
NGC 1377 (IRAS 03344$-$2103) & 0.005 & 0.44 & 1.81 & 7.25 & 5.74 & 43.5
(10.0) \tablenotemark{a} & 0.25 (W) & unknown (iii) \\  
IC 860 (IRAS 13126+2452) & 0.013 & $<$0.09 & 1.27 & 17.9 & 18.1 & 44.6
(11.0) & 0.07 (C) & unknown (iii) \\  
CGCG 1510.8+0725 (Zw 049.057) & 0.013 & $<$0.09 & 0.77 & 20.8 &
29.4 & 44.7 (11.1) & 0.04 (C) & HII (iii) \\  
Cygnus A & 0.056 & $<$0.25 & 1.06 & 2.33 & $<$8.28 & 45.1--45.4
(11.6--11.9) & 0.46 (W) & Sy2 (iv) \\  
3C 234 & 0.185 & 0.16 & 0.26 & 0.23 & $<$0.35 & 45.8 (12.2) & 1.12 (W) &
Sy2 \tablenotemark{b} (v,vi,vii) \\  
\enddata

\tablecomments{
Col.(1): Object.
Col.(2): Redshift. 
Col.(3)--(6): f$_{12}$, f$_{25}$, f$_{60}$, and f$_{100}$ are 
{\it IRAS FSC}
fluxes at 12, 25, 60, and 100 $\mu$m, respectively.
Col.(7): Decimal logarithm of infrared (8--1000 $\mu$m) luminosity
in ergs s$^{-1}$ calculated with
$L_{\rm IR} = 2.1 \times 10^{39} \times$ D(Mpc)$^{2}$
$\times$ (13.48 $\times$ $f_{12}$ + 5.16 $\times$ $f_{25}$ +
$2.58 \times f_{60} + f_{100}$) ergs s$^{-1}$
\citep{sam96}. 
The values in parentheses are the decimal logarithms of the infrared
luminosities in units of solar luminosities.
Col.(8): {\it IRAS} 25 $\mu$m to 60 $\mu$m flux ratio (f$_{25}$/f$_{60}$).
Galaxies with f$_{25}$/f$_{60}$ $<$ 0.2 and $>$ 0.2 are
classified as cool and warm (denoted as ``C'' and ``W''), respectively
\citep{san88}.
Col.(9): Optical spectral classification and references.
(i): \citet{cra96}.
(ii): \citet{wu98}.  
(iii): \citet{vei95}.
(iv): \citet{ost75}.
(v): \citet{gra78}. 
(vi): \citet{ant90}.
(vii): \citet{you98}.
}

\tablenotetext{a}{
The infrared luminosity of NGC 1377 is lower than L$_{\rm IR}$ $\sim$ 
10$^{11}$L$_{\odot}$, but this source is included because there is
a strong buried AGN signature ($\S$ 2.2).
}

\tablenotetext{b}{
3C 234 shows a broad component of optical H$\alpha$ 
emission \citep{gra78}, but it is more likely to be a scattered
component than a direct component \citep{ant90,you98}.
We optically classify this source as a Seyfert 2.}

\end{deluxetable}

%%%%%%%%%%  Table 2 %%%%%%%%%%
\begin{deluxetable}{llccclccc}
\tabletypesize{\scriptsize}
\tablecaption{Observing log \label{tbl-2}}
\tablewidth{0pt}
\tablehead{
\colhead{Object} & 
\colhead{Date} & 
\colhead{Telescope} & 
\colhead{Integration} & 
\colhead{P.A.}  & 
\multicolumn{4}{c}{Standard Stars} \\
\colhead{} & 
\colhead{(UT)} & 
\colhead{Instrument} & 
\colhead{(Min)} &
\colhead{($^{\circ}$)} &
\colhead{Name} &  
\colhead{$L$-mag} &  
\colhead{Type} &  
\colhead{$T_{\rm eff}$ (K)}  \\
\colhead{(1)} & \colhead{(2)} & \colhead{(3)} & \colhead{(4)}
& \colhead{(5)} & \colhead{(6)}  & \colhead{(7)} & \colhead{(8)} &
\colhead{(9)}
}
\startdata
IRAS 04154+1755 & 2005 February 21 & Subaru IRCS & 32 & 0 & HR 1061 & 5.1
& A0V & 9480 \\
NGC 828 & 2005 February 22 & Subaru IRCS & 24 & 0 & HR 761 & 4.9 & F8V &
6000 \\
IRAS 15250+3609 & 2005 February 20 & Subaru IRCS & 16 & 0 & HR 5630 & 5.0
& F8V & 6000 \\
IRAS 17208$-$0014 & 2005 February 21 & Subaru IRCS & 24 & 0 & HR 6439 &
4.4 & G9V & 5400 \\ 
NGC 1377 & 2005 February 20 & Subaru IRCS & 32 & 0 & HR 1532 & 4.0 &
G2.5V & 5800 \\
IC 860 & 2005 April 28 & IRTF SpeX & 92 & 0 & HR 4864 & 4.5 & G7V & 5400 \\
CGCG 1510.8+0725 & 2005 April 30 & IRTF SpeX & 100 & 0 & HR 5567 & 5.8 &
A0V & 9480 \\  
Cygnus A & 2005 May 29 & Subaru IRCS & 24 & 0 & HR 7504 & 4.7 & G2.5V &
5800 \\ 
3C 234 & 2005 February 23 & Subaru IRCS & 24 & 0 & HR 3815 & 3.6 & G8V &
5400 \\ 
\enddata

\tablecomments{Column (1): Object name.
Col. (2): Observing date in UT.
Col. (3): Telescope and instrument. 
Col. (4): Net on-source integration time in min.
Col. (5): Position angle of the slit.
          P.A.(Position angle) = 0$^{\circ}$ corresponds to the
          north-south direction.
Col. (6): Standard star name.
Col. (7): Adopted $L$-band magnitude.
Col. (8): Stellar spectral type.
Col. (9): Effective temperature.}

\end{deluxetable}

%%%%%%%%%%  Table 3 %%%%%%%%%%
\begin{deluxetable}{lcccc}
\tabletypesize{\scriptsize}
\tablecaption{Properties of the 3.3 $\mu$m PAH emission 
\label{tbl-3}}
\tablewidth{0pt}
\tablehead{
\colhead{Object} & 
\colhead{f$_{3.3 \rm PAH}$} & 
\colhead{L$_{3.3 \rm PAH}$}  & 
\colhead{L$_{3.3 \rm PAH}$/L$_{\rm IR}$} & 
\colhead{rest EW$_{3.3 \rm PAH}$} \\
\colhead{} & 
\colhead{($\times$10$^{-14}$ ergs s$^{-1}$ cm$^{-2}$)} & 
\colhead{($\times$10$^{41}$ergs s$^{-1}$)}  & 
\colhead{($\times$10$^{-3}$)} & 
\colhead{(nm)} \\
\colhead{(1)} & \colhead{(2)} & \colhead{(3)} & \colhead{(4)} & 
\colhead{(5)}  
}
\startdata
IRAS 04154+1755 & 8.0 & 5.0 & 0.2 & 60 \\
NGC 828 & 2.0 \tablenotemark{a} & 0.10 & 0.02 & 20 \\  
IRAS 15250+3609 & 5.5 &3.5 & 0.09 & 110 \\   
IRAS 17208$-$0014 & 9.0 & 3.0 & 0.04 & 80 \tablenotemark{b} \\ 
NGC 1377 & $<$1.1 \tablenotemark{c} & $<$0.005 & $<$0.02 & $<$6 \\  
IC 860  & 3.5 & 0.10 & 0.03 & 40 \\ 
CGCG 1510.8+0725 & 7.8 & 0.25 & 0.05 & 110 \\ 
Cygnus A & $<$0.60 & $<$0.4 & $<$0.03 & $<$6 \\  
3C 234 & $<$2.5 & $<$20 & $<$0.3 & $<$4 \\  
\enddata

\tablecomments{
Col.(1): Object name. 
Col.(2): Observed flux of the 3.3 $\mu$m PAH emission. 
Col.(3): Observed luminosity of the 3.3 $\mu$m PAH emission.  
Col.(4): Observed 3.3 $\mu$m PAH-to-infrared luminosity ratio. 
Col.(5): Rest-frame equivalent width of the 3.3 $\mu$m PAH emission.  
         Starbursts have EW$\sim$100 nm (see text).
}

\tablenotetext{a}{The observed PAH profile in this source is slightly
wider than the adopted template spectral shape.
However, the resulting uncertainty of the PAH flux is insignificant. 
We employ the same spectral template as other sources for consistency.
} 

\tablenotetext{b}{This value is similar to that recently estimated by
\citet{ris06} using independent data.} 

\tablenotetext{c}{
The estimated upper limit is a conservative one, by assuming the
lowest plausible continuum level. 
However, we interpret that an apparent excess at 
$\lambda_{\rm rest}$ $\sim$ 3.3 $\mu$m is mostly ascribed to the
artifact of the 3.4 $\mu$m bare carbonaceous dust absorption feature
located at the longer wavelength side. 
} 

\end{deluxetable}

\clearpage

%--------------  Figure 1 -----------------%
\begin{figure}
\includegraphics[angle=-90,scale=.35]{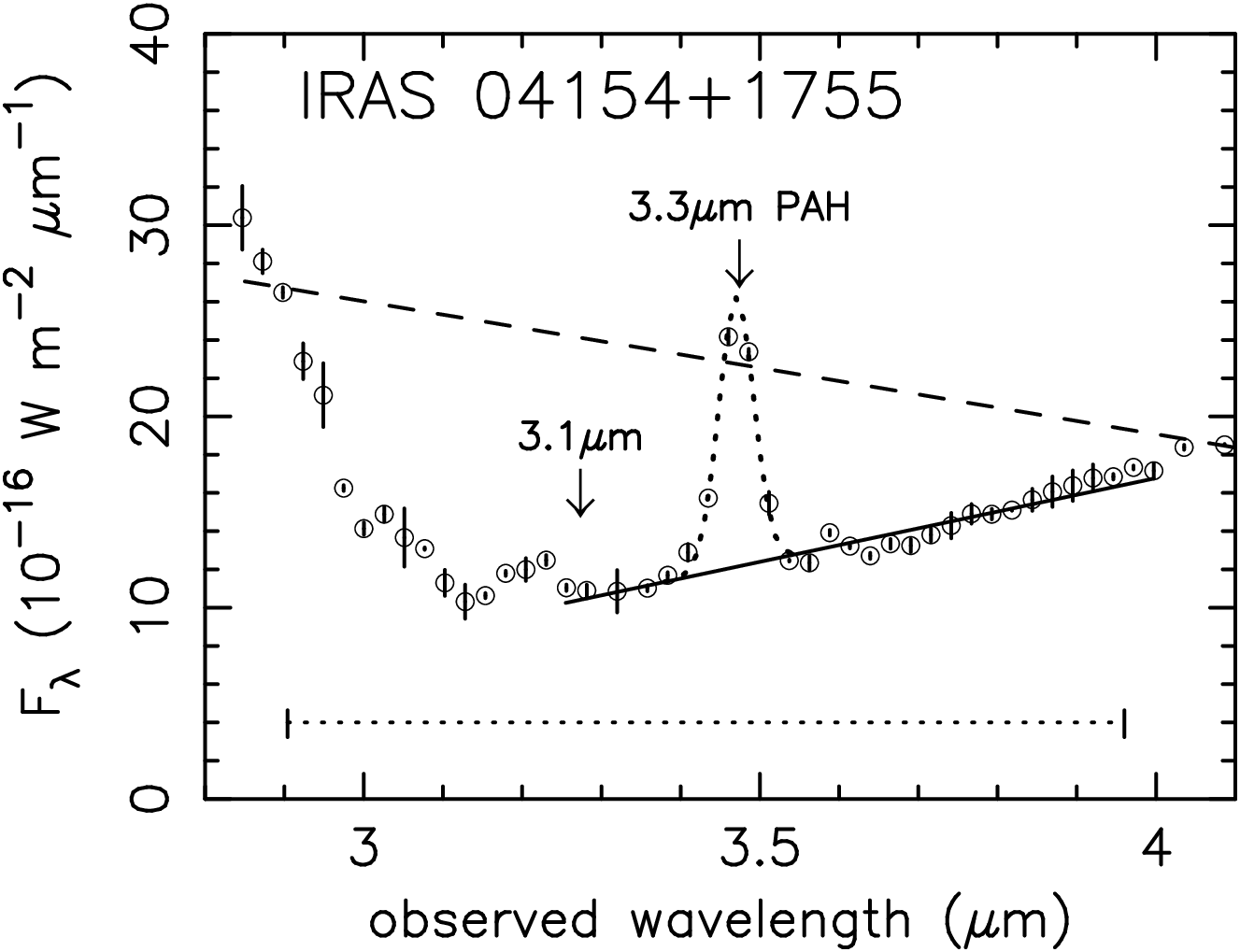} \hspace{0.3cm}
\includegraphics[angle=-90,scale=.35]{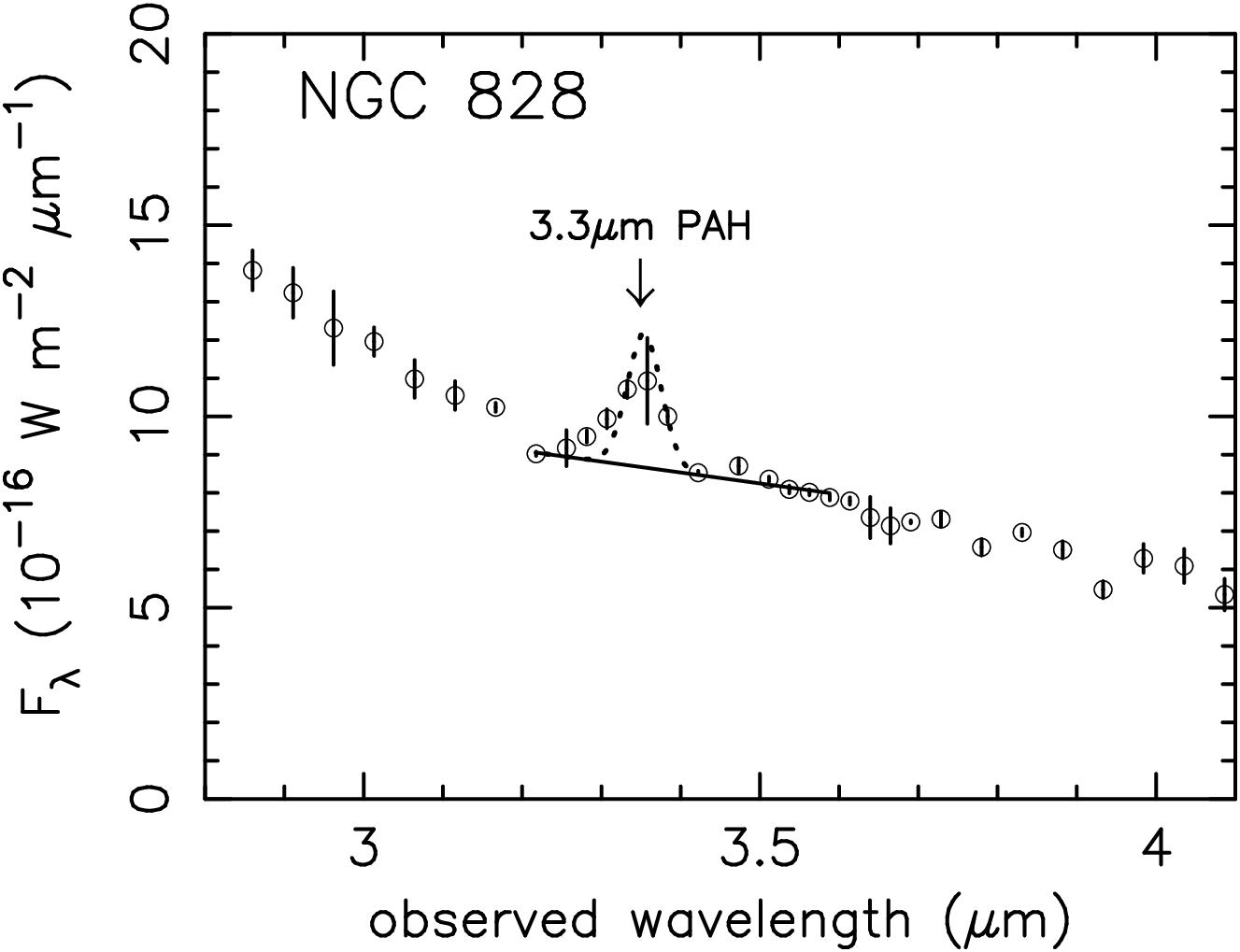} \\ 
\includegraphics[angle=-90,scale=.35]{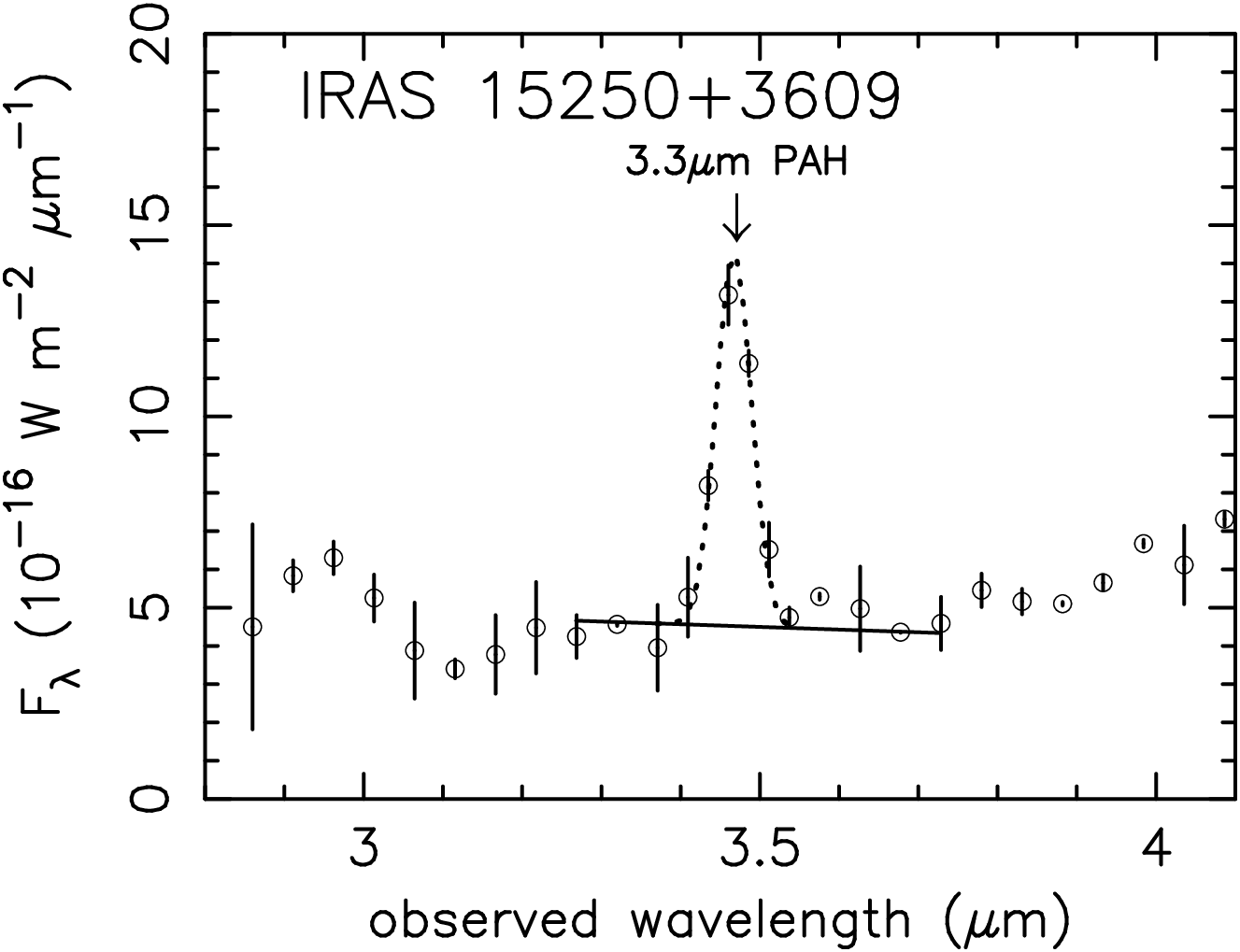} \hspace{0.3cm}
\includegraphics[angle=-90,scale=.35]{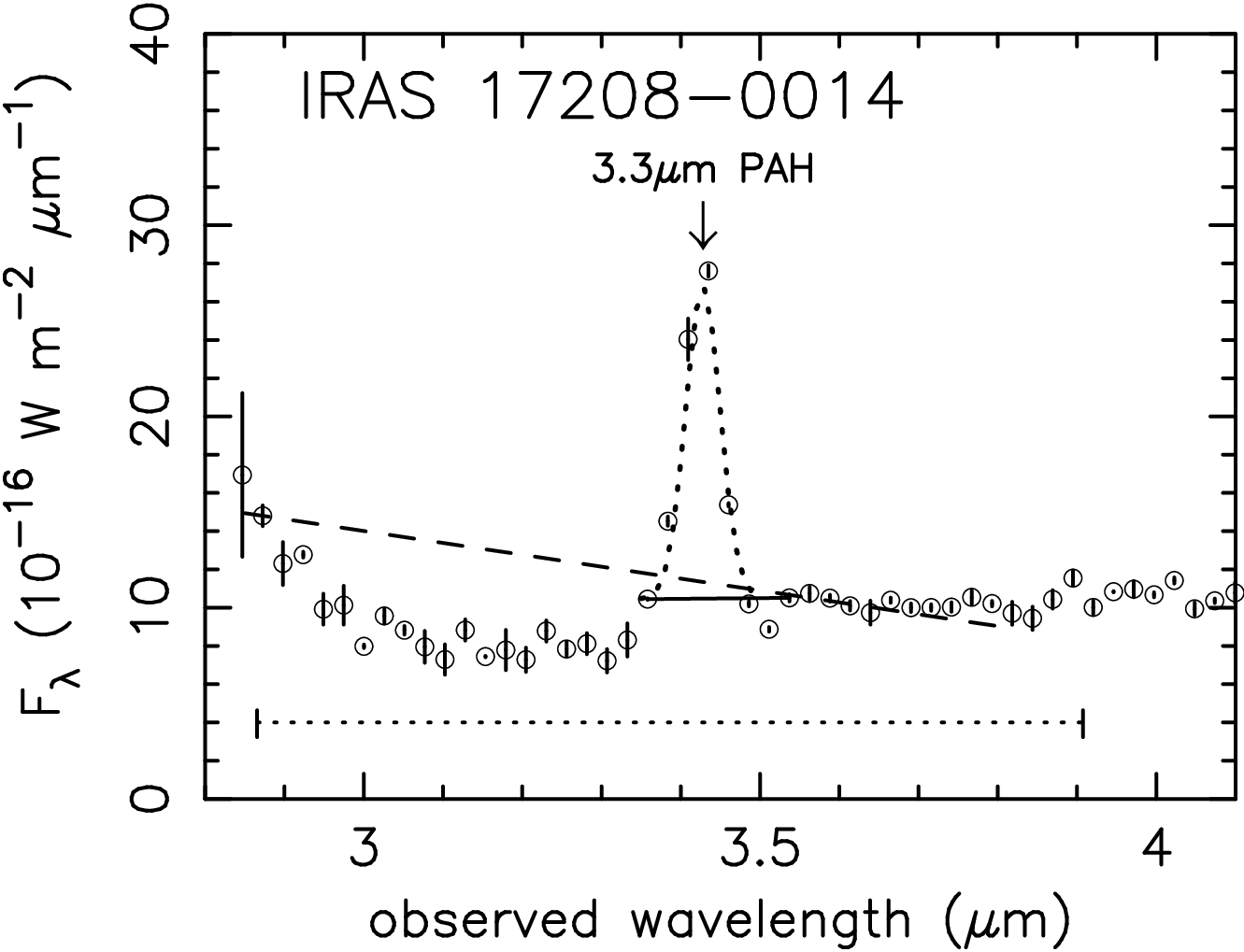} \\ 
\includegraphics[angle=-90,scale=.35]{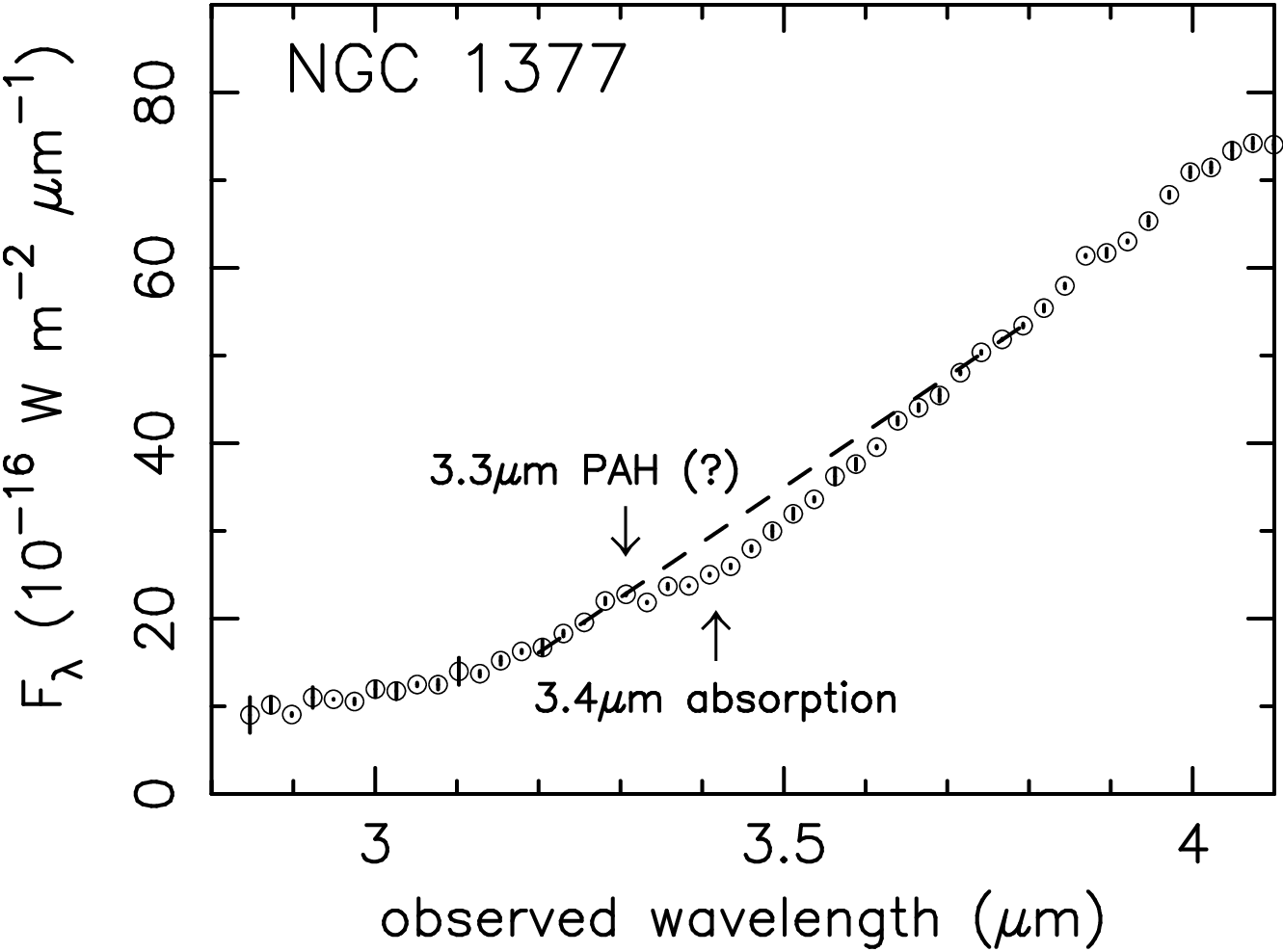} \hspace{0.3cm}
\includegraphics[angle=-90,scale=.35]{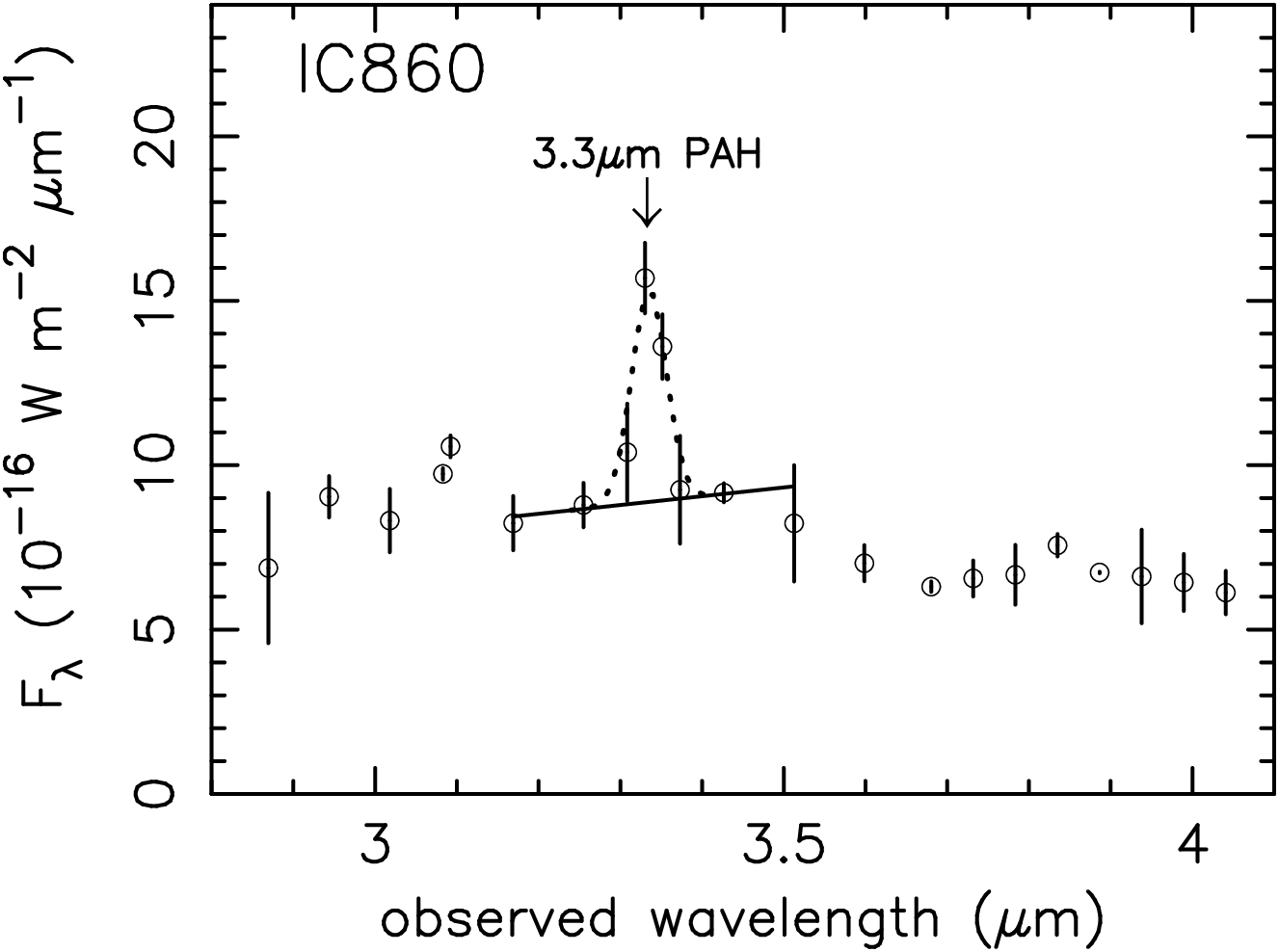} \\
\end{figure}

\clearpage
\begin{figure}
\includegraphics[angle=-90,scale=.35]{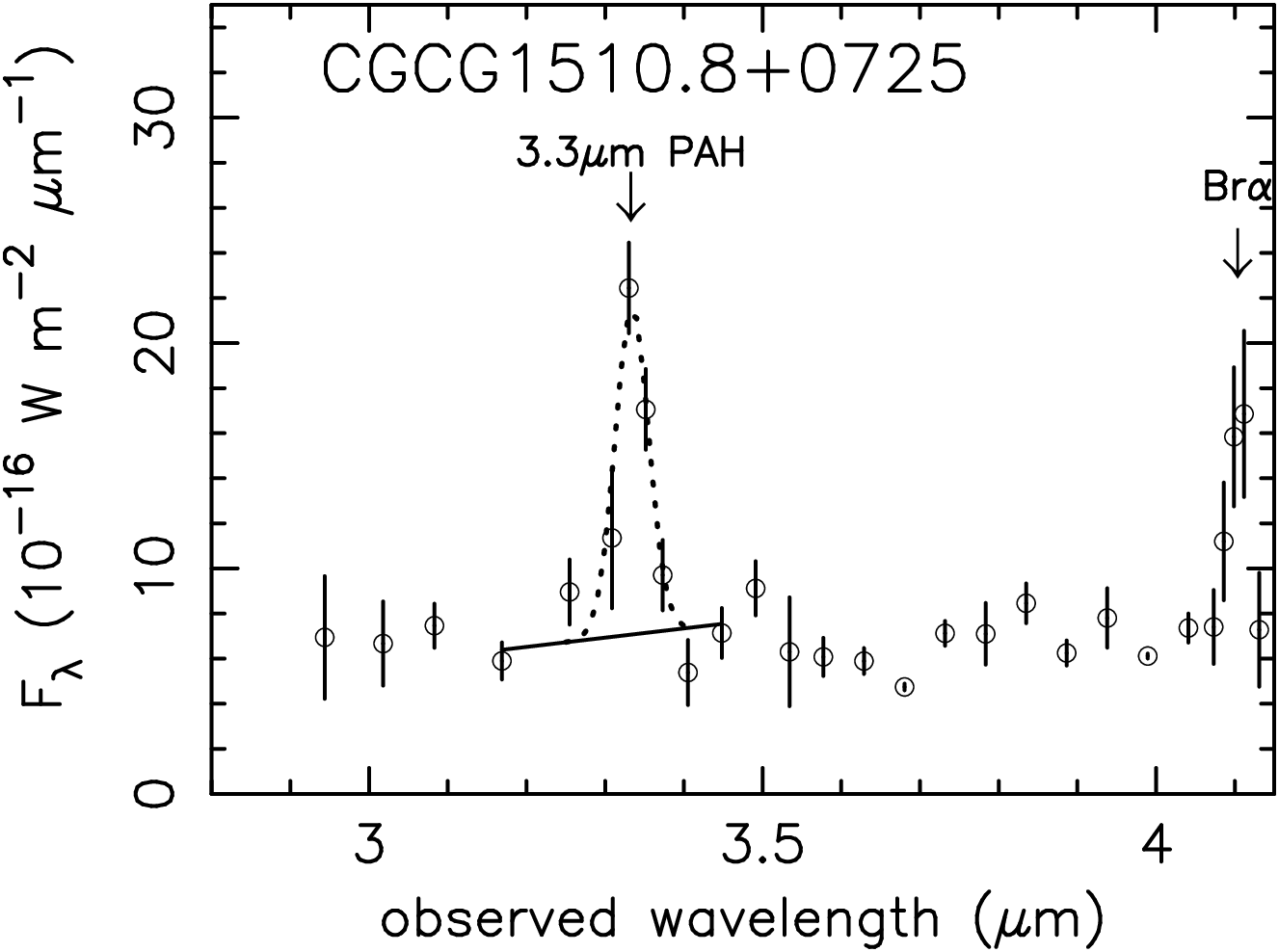} \hspace{0.3cm}  
\includegraphics[angle=-90,scale=.35]{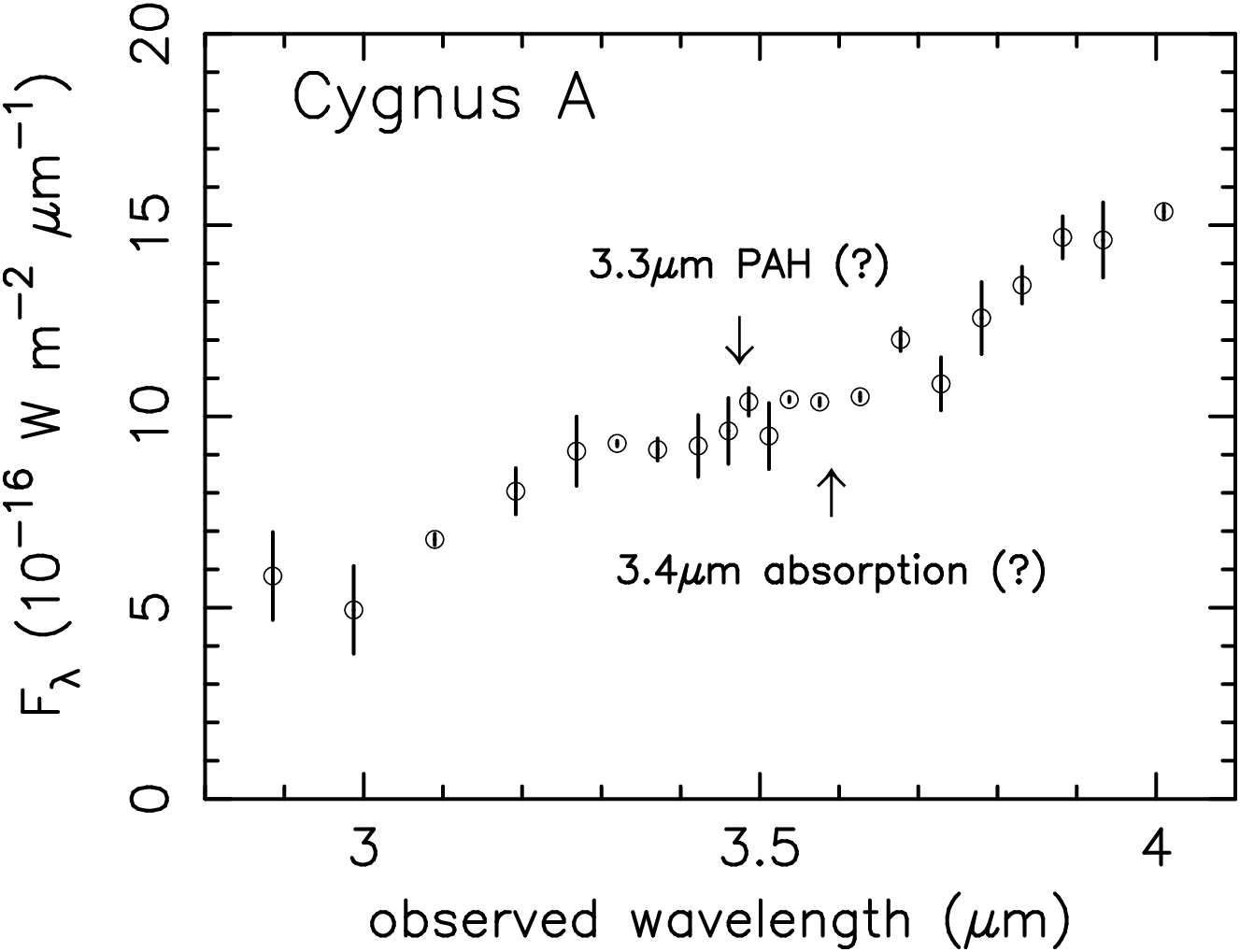} \\
\includegraphics[angle=-90,scale=.35]{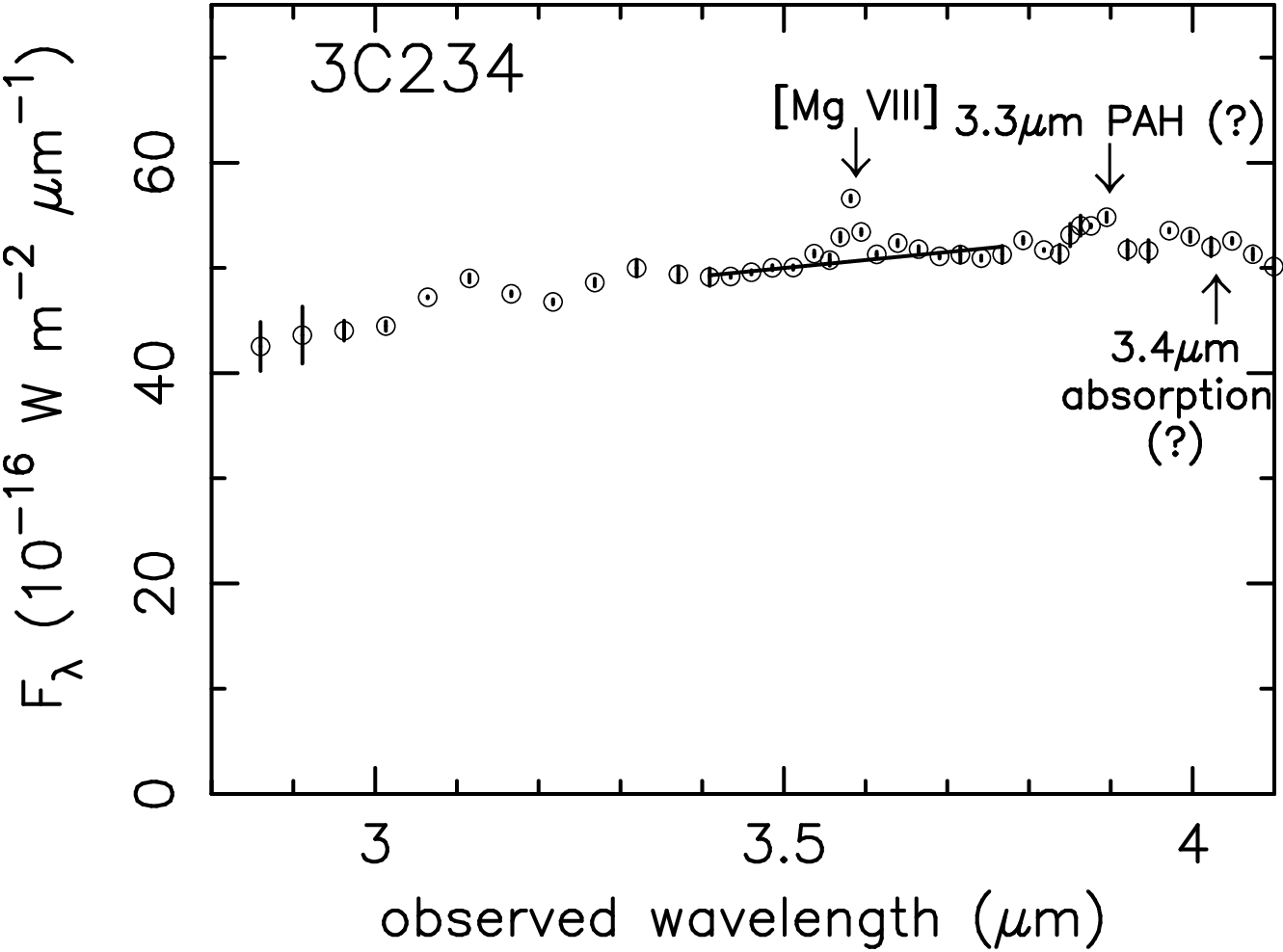} \hspace{0.3cm}  
\caption{
Infrared $L$-band (2.8--4.1 $\mu$m) spectra of the observed
nine infrared luminous galaxies.  The abscissa and ordinate are the
observed wavelength in $\mu$m and flux F$_{\lambda}$ in 10$^{-16}$ W
m$^{-2}$ $\mu$m$^{-1}$, respectively.  
For CGCG 1510.8+0725, the
spectrum up to $\lambda_{\rm obs}$ = 4.15 $\mu$m is presented to show
the detected strong Br$\alpha$ emission line.  
For sources with clearly detectable PAH emission, 
dotted lines indicate the fittings of the PAH emission, using the
template profile ($\S$ 4.1.1), and solid lines are the adopted continuum
levels to estimate 3.3 $\mu$m PAH fluxes.   
The dashed lines in the IRAS 04154+1755, IRAS 17208$-$0014, and NGC 1377
spectra are adopted continuum levels, which are used to measure the
optical depths of 3.1 $\mu$m ice-covered (IRAS 04154+1755 and IRAS
17208$-$0014) or 3.4 $\mu$m bare carbonaceous dust absorption features
(NGC 1377).  In the 3C 234 spectrum, the continuum level used to 
measure the [MgVIII] line flux is shown as a solid line. The horizontal
dotted straight lines, inserted between the two short vertical solid
lines, at the lower part of the spectra of IRAS 04154+1755 and IRAS
17208$-$0014 indicate the wavelength range where the effects of the
broad 3.1 $\mu$m ice absorption feature can be significant
($\lambda_{\rm rest}$ = 2.75--3.75 $\mu$m adopted from the spectrum of
GL 2591; Smith et al. 1989). 
}
\end{figure}

\clearpage

%--------------  Figure 2 -----------------%
\begin{figure}
\includegraphics[angle=-90,scale=.35]{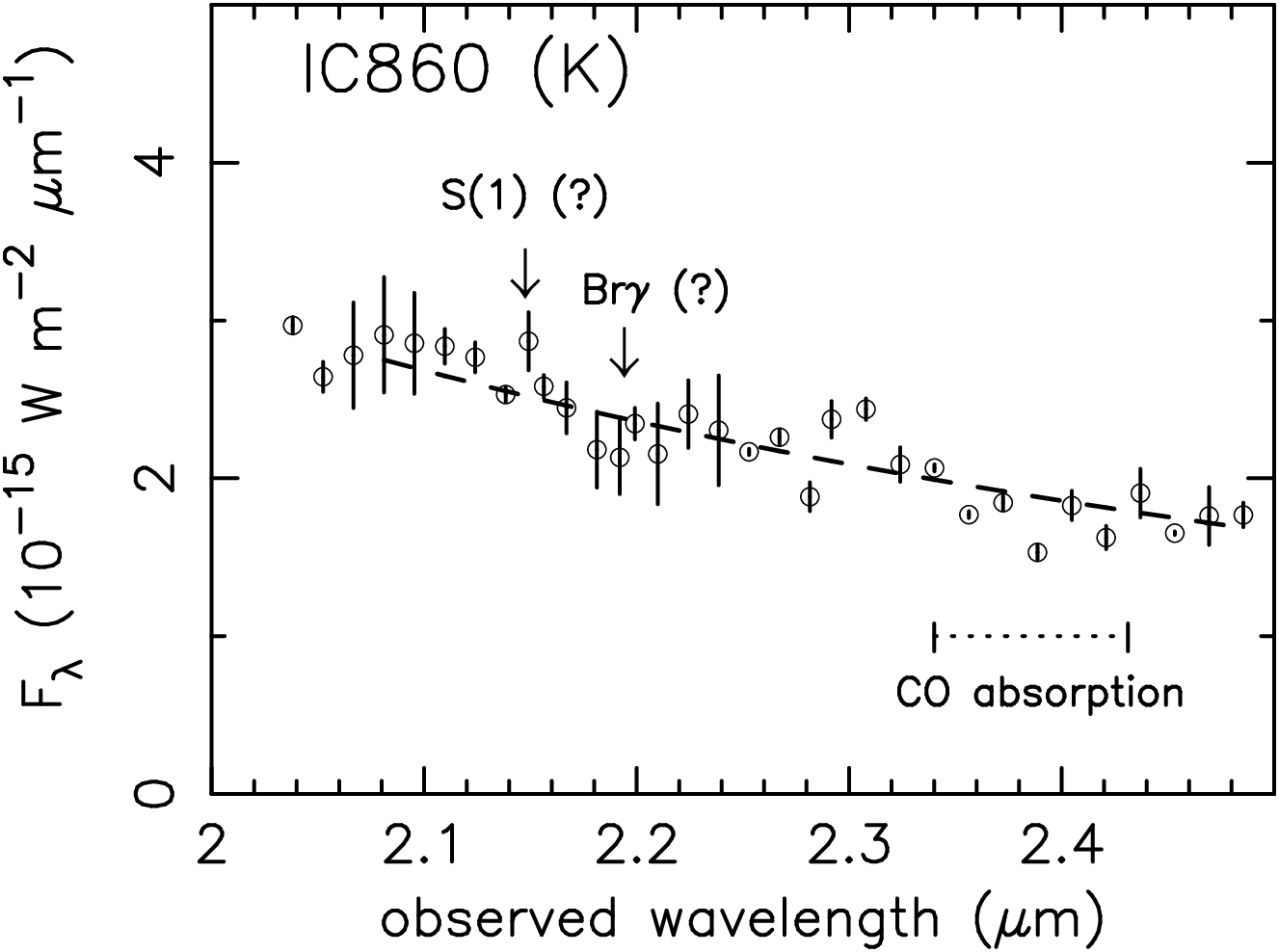} \hspace{0.3cm}
\includegraphics[angle=-90,scale=.35]{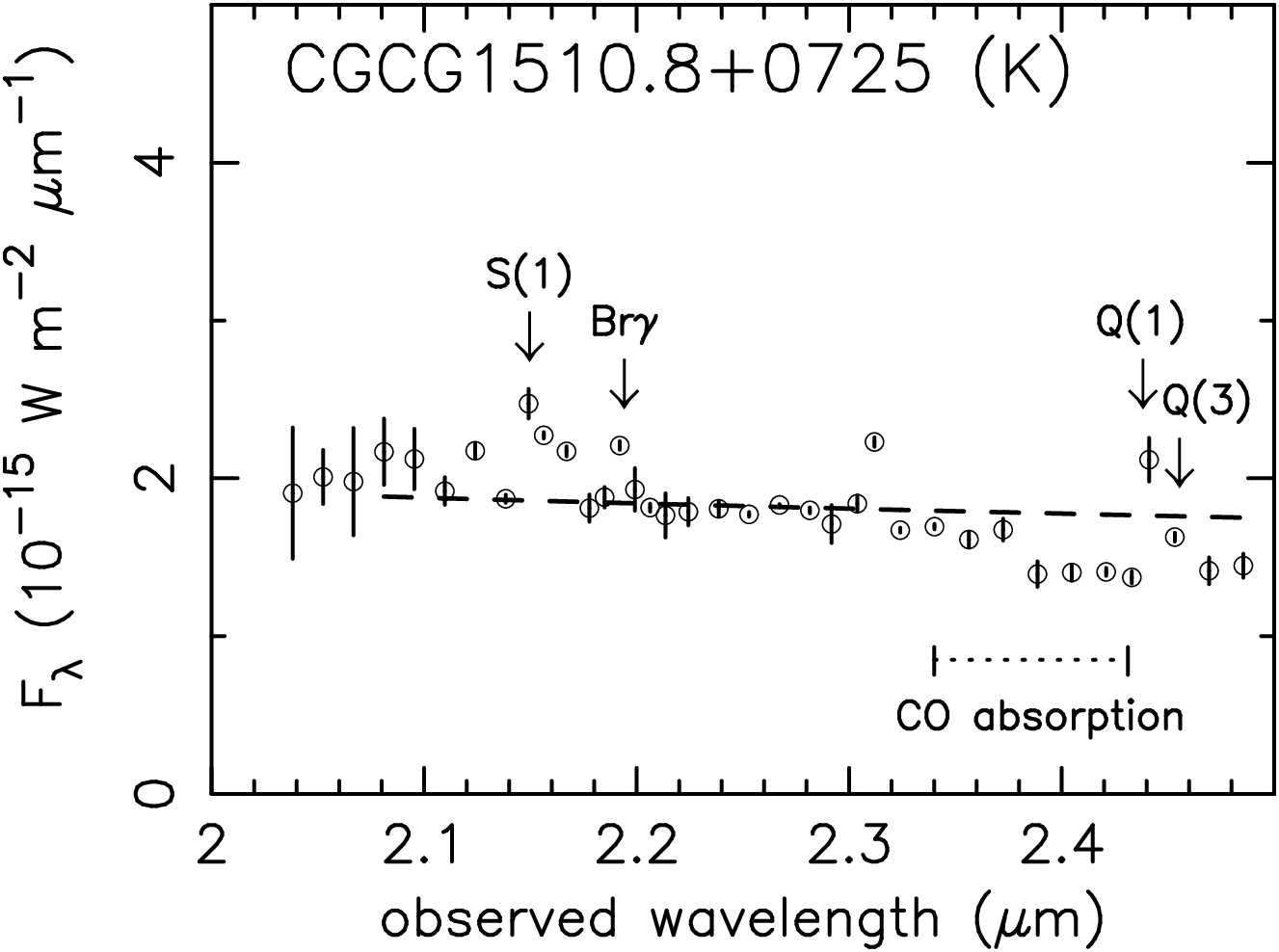}  \\
\caption{
Infrared $K$-band (2.0--2.5 $\mu$m) spectra of two sources
observed with IRTF SpeX (IC 860 and CGCG 1510.8+0725).  The abscissa
and ordinate are the observed wavelength in $\mu$m and flux
F$_{\lambda}$ in 10$^{-15}$ W m$^{-2}$ $\mu$m$^{-1}$, respectively.
The dotted lines inserted between the two vertical lines indicate the
wavelength range of the CO absorption feature at $\lambda_{\rm rest}$
= 2.31--2.40 $\mu$m.  The dashed lines are the continuum levels
adopted to measure the strengths of the CO absorption feature.  Some
detected emission lines are indicated.  S(1): H$_{2}$ 1--0 S(1) at
$\lambda_{\rm rest}$ = 2.122 $\mu$m.  Q(1): H$_{2}$ 1--0 Q(1) at
$\lambda_{\rm rest}$ = 2.407 $\mu$m.  Q(3): H$_{2}$ 1--0 Q(3) at
$\lambda_{\rm rest}$ = 2.424 $\mu$m.  Br$\gamma$: Br$\gamma$ emission
line at $\lambda_{\rm rest}$ = 2.166 $\mu$m.
}
\end{figure}

%\clearpage

%--------------  Figure 3 -----------------%
\begin{figure}
\includegraphics[angle=-90,scale=.35]{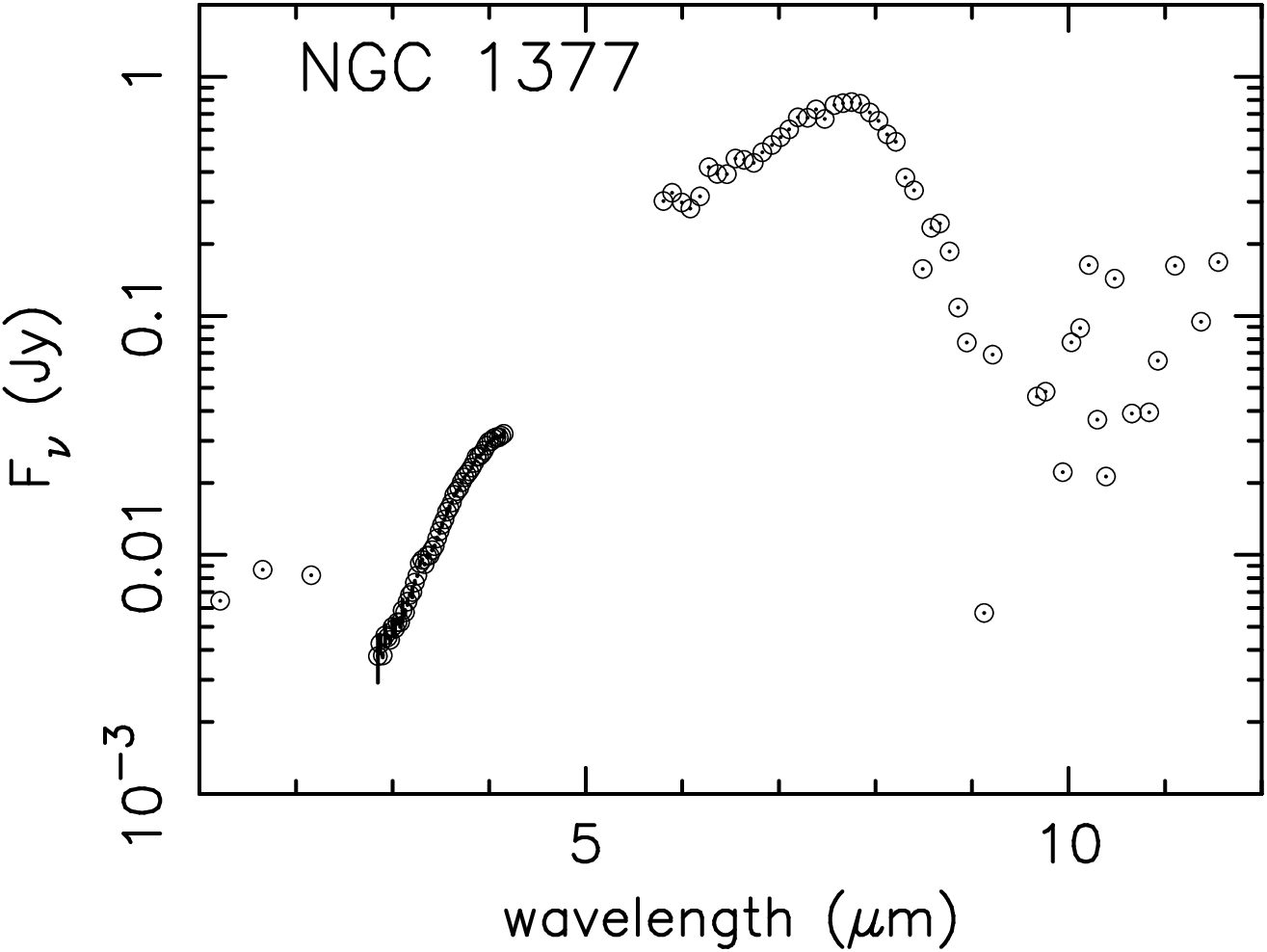}  \hspace{0.3cm}
\includegraphics[angle=-90,scale=.35]{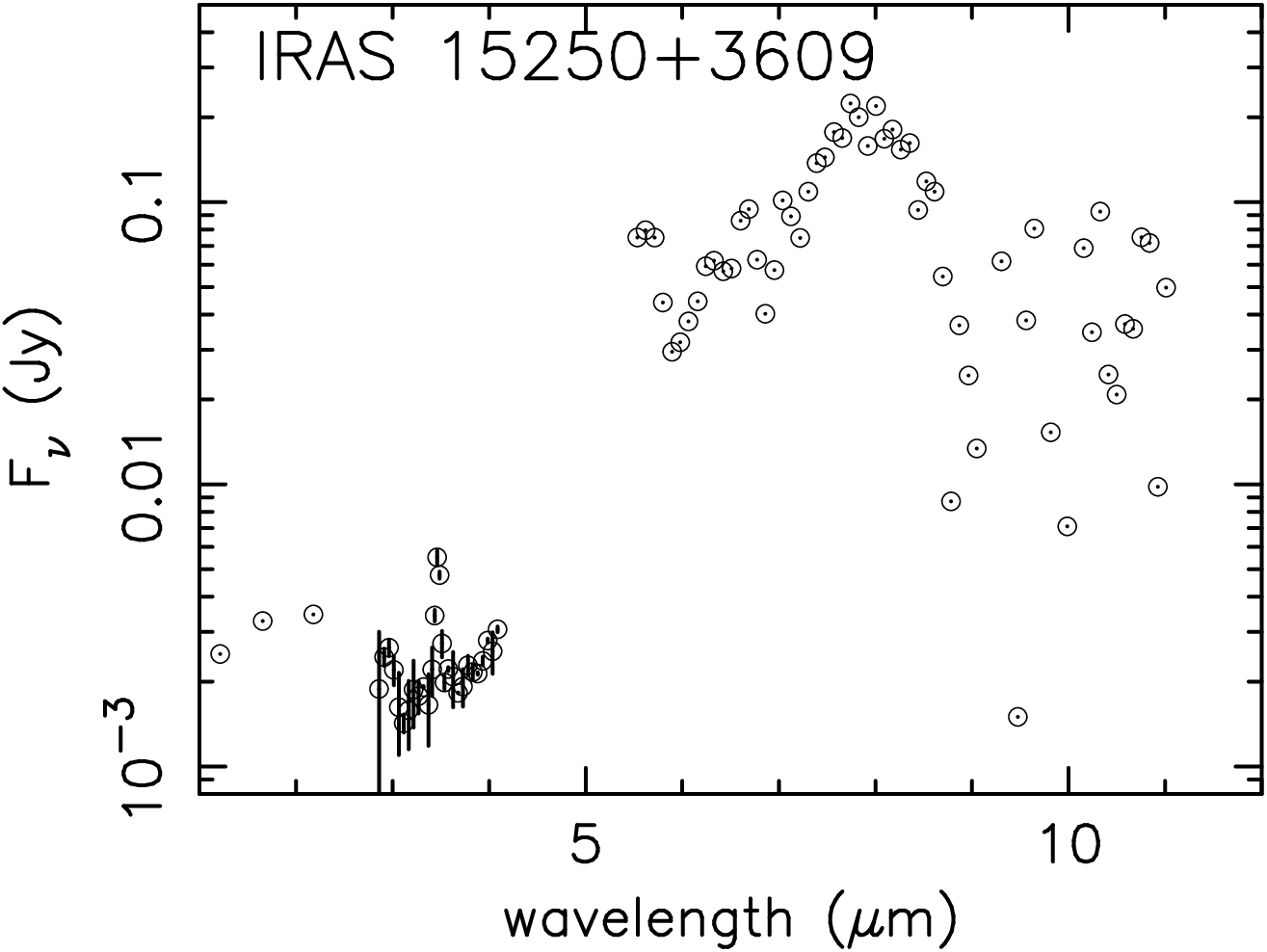} \\
\caption{
Infrared 1--11.5 $\mu$m spectral energy distributions of NGC 1377 
({\it Left}) and IRAS 15250+3609 ({\it Right}).
The abscissa and ordinate are the observed wavelength in $\mu$m and
flux F$_{\nu}$ in Jy, respectively.
Both sources show absorption-dominated 5--11.5 $\mu$m mid-infrared
spectra, with very weak PAH emission, but the $L$-band spectral shapes are
substantially different.
Data points are from 2MASS (1--2.5 $\mu$m), our $L$-band spectra
(2.8--4.1 $\mu$m), and {\it ISO} spectra (5--11.5 $\mu$m). 
}
\end{figure}

\end{document}